\pdfoutput=1

\documentclass{ssdbm2}

\usepackage{graphicx}
\usepackage{subfigure}
\usepackage{url}
\usepackage{amsmath}
\usepackage{amssymb}
\usepackage{units}
\usepackage[lined,boxed,commentsnumbered]{algorithm2e}
\usepackage[usenames,dvipsnames]{color}

\graphicspath{{./figs/}}

\newcommand{\nop}[1]{}
\newcommand{\hide}[1]{}

\newtheorem{definition}{Definition}[section]

\newtheorem{problem}{Problem}[section]

\begin{document}

\title{Node Classification in Uncertain Graphs}

\numberofauthors{3}

\author{
\alignauthor
Michele Dallachiesa\\
       \affaddr{University of Trento, Italy}\\
       \email{dallachiesa@disi.unitn.it}
\alignauthor
Charu Aggarwal\\
       \affaddr{IBM T.J. Watson Research~Center}\\
       \email{charu@us.ibm.com}
\alignauthor
Themis Palpanas\\
       \affaddr{Paris Descartes University}\\
       \email{themis@mi.parisdescartes.fr}
}

\maketitle

\begin{abstract}
In many real applications that use and analyze networked data, the
links in the network graph may be erroneous, or derived from
probabilistic techniques. In such cases, the node
classification problem  can be challenging, since the unreliability
of the links may affect the final results of the classification
process. If the information about link reliability is not used
explicitly, the classification accuracy in the underlying network
may be affected adversely. 
In this paper, we focus on situations that require the analysis of the uncertainty that is present in the graph structure. 
We study the novel problem of node classification in uncertain graphs, by treating uncertainty as a first-class citizen. 
We propose two techniques based on a Bayes model and automatic parameter selection, and show that the incorporation of uncertainty in the classification process as a first-class citizen is beneficial. 
We experimentally evaluate the proposed approach using different real data sets, and study the behavior of the algorithms under different conditions. 
The results demonstrate the effectiveness and efficiency of our approach.

\end{abstract}

\hide{
\category{H.4}{Information Systems Applications}{Miscellaneous}
}

\keywords{Network Classification, Structural Classification,
Label Propagation} 

%
%
%
%

\section{Introduction}
The problem of collective classification is a widely studied one  in the
context of graph mining and social networking applications. In this
problem, we have a network containing nodes and edges, which can be represented as a graph. 
Nodes in this network may be labeled, but it is not necessary that all nodes have a label.
Typically, such labels may represent some properties of interest in the underlying network.
This is a setting that appears in several situations in practice.

Some examples of such labeled networks in real scenarios are listed
below:
\begin{itemize} \item  In a bibliographic network, nodes correspond
to authors, and the edges between them correspond to co-authorship
links.  The labels in the bibliographic network may correspond to
subject areas that experts are interested in. It is desirable to use
this information in order to classify other nodes in the network.
\item  In a biological network, the  nodes correspond to the proteins. The edges
may represent the possibility that the proteins  may interact. The
labels may correspond to properties of proteins \cite{eronen2012biomine}.
\item In a movie-actor network, the nodes correspond to the actors. The
edges correspond to the co-actor relationship between the different
actors.  The labels  correspond to the pre-dominant genre of the
movie of the actor.
\item In a patent network, the nodes correspond to patent assignees.
The edges model the citations between the respective patents.
The labels correspond to the class categories.
\end{itemize}

In such networks, only a small fraction of the nodes may be labeled,
and these labels may be used in order to determine the labels of
other nodes in the network. This problem is popularly referred to as
{\em collective classification} or {\em label propagation}
\cite{agg10,cormode07,Bilgic08,getoor03,ji2011ranking,Zhou03,Zhou05,Zhu03},
and a wide variety of methods have been proposed for this problem.

The problem of data uncertainty has been widely studied in the
database literature \cite{ubook,DallachiesaNMP12,DallachiesaVLDB15}, and also presents numerous
challenges in the context of network data \cite{ren2009naive}.
In many real networks,
the links\footnote{In the rest of this paper we use the terms \emph{network} and \emph{graph}, as well as \emph{link} and \emph{edge}, interchangeably.} are uncertain in nature, and are derived with the use of a
probabilistic process. In such cases, a probability value may be
associated with each edge. Some examples are as follows:
\begin{itemize}
\item In biological networks, the links are derived from
probabilistic processes. In such cases, the edges have uncertainty
associated with them. Nevertheless, such probabilistic networks are
valuable, since the probability information on the links provides
important information for the mining process.
\item  The links in many military networks are constantly changing
and may be uncertain in nature.  In such cases, the analysis needs
to be performed with imperfect knowledge about the network.
\item  Networks in which some links have large failure probabilities
are uncertain in nature.
\item Many human interaction networks  can be created from real
interaction processes, and such links are often uncertain in
networks.
\end{itemize}
Thus, such networks can be represented as probabilistic networks, in
which we have probabilities associated with the existence of  links.
Such probabilities can be very useful for improving the
effectiveness of problems such as collective classification. 
Furthermore, these networks may also have properties associated
with nodes, that are denoted by labels. 

Recent
years have seen the emergence of numerous  methods for uncertain
graph management \cite{hua,jin,potamias} and mining
{\cite{jin,kong,kollios,lin,papa,zou,zou2}, in which uncertainty is
used directly as a first-class citizen. However, none of these
methods address the problem of collective graph classification.

One possibility is to use {\em sampling} of possible worlds on the edges in order to generate
different instantiations of the underlying network. The collective
classification problem can be solved on these different
instantiations, and  voting can be used in order to report the final
class label. The major disadvantage with this approach is that the
sampling process could result in a sparse or disconnected network
which is not suited to the collective classification problem. In
such cases, good class labels cannot be easily produced with a
modest number of samples.

In this paper, we investigate
the problem of collective classification in  uncertain networks with
a more direct use of the uncertainty information in the network \footnote{A preliminary version of this work appeared in \cite{DallachiesaSSDBM14}.}. We
design two algorithms for collective classification. The first
algorithm uses a probabilistic approach, which explicitly accounts
for the uncertainty in the links in the classification. 

The second
algorithm works with the assumption that most of the information in
the network is encoded in high-probability links, and low-probability
links sometimes even degrade the quality. Therefore, the algorithm uses the links with high
probability in earlier iterations, and successively relaxes the
constraints on the quality of the underlying links. The idea is that
a greater caution in early phases of the algorithm ensures
convergence to a better optimum.

The contributions we make in this paper can be summarized as follows.

\begin{itemize}
\item
We introduce the problem of collective classification in uncertain
graphs, where uncertainty is associated with the edges of the graph,
and provide a formal definition for this problem. 
\item
We introduce two algorithms based on iterative probabilistic labeling that
 incorporate the uncertainty of edges in their operation.
These algorithms are based on a Bayes formulation, which enables them to capture correlations
 across different classes, leading to improved accuracy.
 
\item
We perform an extensive experimental evaluation, using two real datasets from diverse domains.
We evaluate our techniques using a multitude of different conditions, and input data characteristics.
The results demonstrate the effectiveness of the proposed techniques and serve as guidelines for the
practitioners in the field.
\end{itemize}

This paper is organized as follows. In Section~\ref{sec:relwork} we survey prior studies on
collective classification and on mining uncertain networks.
In Section~\ref{sec:cc}, we formally define the problem of collective classification in uncertain networks. In Section~\ref{sec:proposal}, we present our model and two algorithms for collective classification.
We discuss the space and time complexity of our proposal in Section~\ref{sec:complexity}, and we present the results of our experimental evaluation in Section~\ref{sec:results}. 
Finally, we discuss the conclusions in Section~\ref{sec:conclusions}.

%
%
%
%

\section{Related Work}
\label{sec:relwork}

The problem of node classification has been studied in the
graph mining literature, and especially relational data in the
context of {\em label or belief propagation}
\cite{Tasker02,Zhou03,Zhou05}. Such propagation techniques are also
used as a tool for semi-supervised learning with both labeled and
unlabeled examples \cite{Zhu03}. Collective classification
\cite{macskassy2003simple,getoor03,cormode07} refers to
semi-supervised learning methods that exploit the network structure
and node class labels to improve the classification accuracy. These
techniques are mostly based on the  assumption of homophily in
social networks \cite{blau1977inequality,mcpherson2001birds}:
neighboring nodes tend to belong to the same class.  A technique
has been proposed in \cite{getoor03}, which uses link-based
similarity for node-classification in  directed graphs.  Recently,
collective classification methods have also been used in the context
of blogs \cite{cormode07}.
 In \cite{Bilgic08}, Bilgic et al. discuss the problem of
overcoming the propagation of erroneous labels by asking  the user
for more labels.  A method for  performing  collective
classification} of email speech acts has been proposed by Carvalho
et al. in \cite{cohen05}, exploiting the sequential correlation of
emails. In \cite{ji2011ranking}, Ji et al. integrate the
classification of nodes in heterogeneous networks with ranking.
Methods for leveraging label consistency  for collective
classification have been proposed in \cite{Zhou03,Zhou05,Zhu03}.

Recently, the database and data mining community has investigated
the problem of uncertain data mining widely \cite{ubook,DalviS07}. 

A comprehensive review of the proposed models and algorithms
can be found in \cite{AggarwalY09}.
Several database systems supporting uncertain data have been proposed, such as Conquer
\cite{fuxman2005conquer}, Trio \cite{AgrawalBSHNSW06},  MistiQ \cite{DalviS04}, MayMBS
\cite{AntovaKO07} and Orion \cite{SinghMMPHS08}.  

The "possible worlds" model, introduced by Abiteboul et al. \cite{AbiteboulKG87}, formalizes
uncertainty by defining the space of the possible instantiations of the database. Instantiations
must be consistent with the semantics of the data. For example, in a graph database representing
moving object trajectories there may be be different configurations of the edges where each
node represents a region in the space. However, an edge cannot connect a pair of nodes that
represent a pair of non-neighboring regions.
The main advantage of the "possible worlds" model is that the formulations of the queries
originally designed to cope with certain data can be directly applied on each possible instantiation.
Many different alternatives have then been propose to aggregate the results across the
different instantiations.

Despite its attractiveness, the number of possible worlds explodes very quickly
and even their enumeration becomes intractable problem. To overcome these issues, 
simplifying assumptions have been introduced to leverage its simplicity: The 
tuple- and the attribute-uncertainty models \cite{JestesCLY11,AggarwalY09}.
In the attribute-uncertainty model, the uncertain tuple is represented by means of multiple
samples drawn from its Probability Density Function (PDF). In contrast, in the tuple-uncertainty
model the value of the tuple is fixed but the tuple itself may not exist.

Similar simplifications have been considered for graph databases where nodes may or may
not exist (node-uncertainty) and edges are associated with an existence probability (edge-uncertainty).
The underlying uncertainty model can then be used to generate graph instances, eventually considering
additional generation rules to consider correlations across different nodes and edges.
In this study we combine a Bayes approach and the edge-uncertainty model.

The problem of uncertain graph mining has also been
investigated extensively. The  most common problems studied in
uncertain graph management are those of nearest neighbor query
processing \cite{hua,potamias},  reachability computation
\cite{jin2} and subgraph search \cite{yuan}. In the context of
uncertain graph {\em mining}, the problems commonly studied are
frequent subgraph mining \cite{papa,zou,zou2}, reliable subgraph
mining \cite{jin}, and clustering \cite{kollios,lin}. Recently, the
problem of graph classification has also been studied for the
uncertain scenario \cite{kong}, though these methods are designed
for classification of many small graphs, in which labels are
attached to the entire graph rather than a node in the graph.
Typical social and web-based scenarios use a different model of
collective classification, in which the labels are attached to nodes
in a single large graph.

In this work, we study the problem of {\em collective
classification} in the context of uncertain networks, where the
underlying links are uncertain. Uncertainty impacts negatively on the
classification accuracy. First, links may connect sub-networks of
very different density, causing the propagation of erroneous labels.
Second, the farthest distance between two nodes tends to be smaller
in very noisy networks, because of the presence of a larger number
of uncertain edges, which include both true and  spurious edges.
This reduces the effectiveness of iterative models because of the faster
propagation of errors. Some of our techniques, which drop
uncertain links at earlier stages of the algorithm,  are designed to
ameliorate these effects.

%
%
%
%

\section{Collective Classification Problem}
\label{sec:cc}

In this section, we  formalize the problem of collective
classification after introducing some definitions. An uncertain
network is composed of nodes whose  connections may exist with some
probability.

\begin{definition}[Uncertain Network]
\label{def:unet}  An uncertain network is denoted by $G=(N, A, P)$,
with node set $N$, edge set $A$ and probability set $P$. Each edge
$(i, j) \in A$ is associated with a probability value $p_{ij} \in
P$. This is the probability that edge $(i, j)$ exists in the
network.
\end{definition}

We assume that the network is undirected, though the method can easily
be extended to the directed scenario.
We can assume that the $|N|\times |N|$ matrix $P$  has entries which
are denoted by $p_{ij}$ and $p_{ij} = p_{ji}$.  A node $i \in N$ can
be associated with a label, representing its membership in a class.
For ease in notation, we assume that node labels are integers.

\begin{definition}[Node Label]
\label{def:label} Given a set of labels $S$ drawn from a set of
integers $\{ 1 \ldots l \}$, we denote the label of node $i$ by
$L(i)$. If a node $i$ is unlabeled, the special label $0$ is used.
\end{definition}

We can now introduce the definition of the collective classification
problem on uncertain graphs.

\begin{problem}[Uncertain Collective Classification]
\label{problem:cc} Given an uncertain network $G=(N, A, P)$ and the
subset of labeled nodes $T_0 = \{ i \in N : L(i) \neq 0 \}$, predict
the labels of nodes in $ N - T_0$.
\end{problem}

\begin{figure}[htb]
\begin{center}
\vspace{0.1cm}
\resizebox{62mm}{!}{\includegraphics[angle=0]{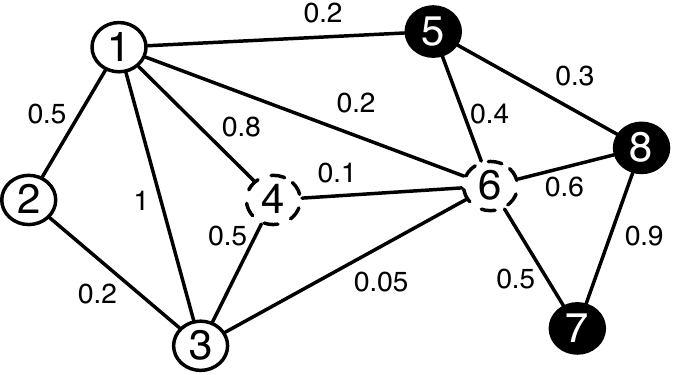}}
\caption{\small\bf Example of uncertain network. Nodes $\{1,2,3\}$ are labeled {\em white} and nodes $\{5,8,7\}$ are labeled {\em black}, while labels for nodes $\{4,6\}$ are unknown. Edges between nodes exist with some probability.}
\label{fig:net1}
\vspace{0.1cm}
\end{center}
\end{figure}

Figure~\ref{fig:net1} shows an example of an uncertain network.
Nodes  $1$, $2$, and $3$ are labeled {\em white}, and nodes $5$, $7$,
and $8$ are labeled {\em black}. The label of nodes $4$ and $6$ is
unknown. The aim of collective classification is to assign labels
to nodes $4$ and $6$.

%
%
%
%

\section{Iterative Probabilistic Labeling}
\label{sec:proposal}

In this section, we first present the algorithm for iterative
probabilistic  labeling. A Bayes approach is used in order to
perform the iterative probabilistic labeling. This method models the
probabilities of the nodes belonging to different classes on the
basis of  the adjacency behavior of the nodes.  The Bayes approach
can directly incorporate the edge uncertainty probabilities into the
estimation process.
We continue with a second algorithm that builds upon the first one,
and is based on iterative edge augmentation.
Finally, we describe a variation of the second algorithm that is a
linear combination of two classifiers.

\subsection{Bayes Approach}

The overall approach for the labeling process  uses a Bayesian model
for the labeling. In the rest of the paper, we refer to this
algorithm as {\em uBayes}.
Given that we have an unlabeled node $r$, which
is adjacent to $s$ other nodes denoted by $t_1 \ldots t_s$, how do
we determine the label of the node $r$?  It should be noted that the
concept of adjacency is also uncertain, because the edges are
associated with probabilities of existence. This is particularly
true, when the  edge probabilities are relatively small, since the
individual network instantiations are likely to be much sparser and
different than the probabilistic descriptions.  Furthermore, for
each edge $(i, j)$ we need to estimate the probability of the node
$j$ having a particular label value, given the current value of the
label at node $i$. This is done with the use of training data
containing the labels and edges in the network. These labels and
edges can be used to construct a Bayesian model of how the labels on
the nodes and edges relate to one another.

The algorithm uses an iterative approach, which successively labels
more nodes in different iterations. This is the set $T$ of nodes
whose labels will not be changed any
further by the algorithm. Initially, the algorithm starts off by
setting $T$ to the initial set of (already) labeled nodes $T_0$.  The set in
$T$ is expanded to $T \cup T^+$ in each iteration, where
$T^+$ is the set of nodes not yet labeled that are adjacent to
the labeled nodes in $T$. If $T^+$ is empty, either all nodes have
been labeled or there is a disconnected component of the network
whose nodes are not in $T_0$.

The expanded
set of labeled nodes are added to the set of training nodes in order
to compute the propagation probabilities on other edges.
Thus, the overall algorithm
iteratively performs the following steps:
\begin{itemize}
\item Estimating the Bayesian probabilities of propagation from
the current set of edges.
\item Computing the probabilities of the labels of the nodes in
$N-T$.
\item Expanding the set of the nodes in $T$, by adding the set of
nodes from $T^+$, whose labels have the highest probability for a
particular class.
\end{itemize}
These steps are repeated until no more nodes reachable
from the set $T$ remain to be labeled. We then label all the remaining
nodes in a single step, and terminate. The overall procedure for
performing the analysis is illustrated in Algorithm~\ref{iterative}. It
now remains to discuss how  the individual steps in Algorithm~\ref{iterative} are performed.\\

\begin{algorithm}
\begin{tabbing}
{\bf Algorithm} {\em uBayes}(Graph: $G$\\
\ \ \ \ \  Uncertainty Prob.: $P$, Initial Labeling: $T_0$ );\\ \\
{\bf begin}\\
\ \ \ \= $T=T_0$;\\
\> {\bf while} (not termination) {\bf do}\\
\> {\bf begin}\\
\>\ \ \ \= Compute edge propagation probabilities;\\
\>\> Compute node label probabilities in $N-T$; \\
\>\> Expand $T$ with $T^+$ nodes;\\
\> {\bf end}\\
 {\bf end}
\end{tabbing}
\caption{Broad Framework for Uncertain Classification.}
\label{iterative}
\end{algorithm}

The two most important steps are the computation of the
edge-propagation probabilities and the expansion of the node labels
with the use of the Bayes approach. For a given edge $(i, j)$ we
estimate $P(L(i)=p| L(j)=q)$. This  is  estimated from the data in
{\em each iteration} by examining the labels of nodes which have
already been decided. Therefore, the training process is
successively refined in each iteration.  Therefore, the value of
$P(L(i)=p|L(j)=q)$ can be estimated by examining those edges for
which one end point contains a label of $q$. Among these edges, we
compute the fraction for which the other end point contains a label
of  $p$.  For example, in the network shown in Figure~\ref{fig:net1}
the probability $P(L(6)=black | L(5)=black)$ is estimated as $(0.3 +
0.9) / (0.3 + 0.9 + 0.2) = 0.85$.
The label of node $6$ is unknown, and it is not considered in the calculation.
Note that this is simply equal to
the probability that both end points of an edge  are {\em black}, if one
of them is {\em black}. Therefore, one can compute the uncertainty
weighted conditional probabilities for this in the training process
of each iteration.

This provides an estimate for the conditional probability. We note that in some
cases, the number of nodes with a  label of either $p$ or $q$ may be
too small for a robust estimation.  The following smoothing
techniques are useful in reducing the effect of ill-conditioned
probabilities:
\begin{itemize}
\item We always add a small value $\delta$ to each probability. This
is similar to Laplacian smoothing and prevents any
probability value from being zero, which would cause problems in a
multiplicative Bayes model.
\item In some cases, the estimation may not be possible when labels
do not exist for either nodes $p$ or $q$. In those cases, we set the
probabilities to their prior values.
\end{itemize}
The prior is defined as the  value of $P(L(i)=p)$, and  is equal to
the fraction of currently labeled nodes with label of $p$. The prior
therefore  defines the default behavior in cases where the adjacency
information cannot be reasonably used in order to obtain a  better
posteriori estimation.

For an unlabeled node $r$, whose neighbors $i_1 \ldots i_s$ have
labels $t_1 \ldots t_s$, we estimate its (unnormalized) probability  by using the
naive Bayes rule over all the adjacent labeled neighbors. This  is therefore
computed as follows:
\begin{eqnarray*}
\\
P(L(r) = p|L(i_1)=t_1 \ldots L(i_s)=t_s) \propto \ \ \ \ \ \ \ \ \ \ \ \ \ \ \ \\
\ \ \ \ \ \ \ \ \ \ \ \  P(L(r)=p) \cdot \prod_k P(L(i_k)= t_k|L(r)=p)
\end{eqnarray*}


Note that the above model incorporates the uncertainty probabilities
directly within the product term of the equation.
We can perform the estimation for each of the different classes separately.  If desired, one
can normalize the probability values to sum to one. However, such a
normalization is not necessary in our case, since the only purpose of the
computation is to determine the highest probability value in order
to assign labels.

%
%
%
%

\subsection{Iterative Edge Augmentation}

The approach mentioned above is not very effective when a large fraction of
the edges are noisy. 
In particular, if many edges have a low probability, this can have a
significant impact on the classification process. 

\begin{figure}[htb]
\begin{center}
\vspace{0.1cm}
\resizebox{72mm}{!}{\includegraphics[angle=0]{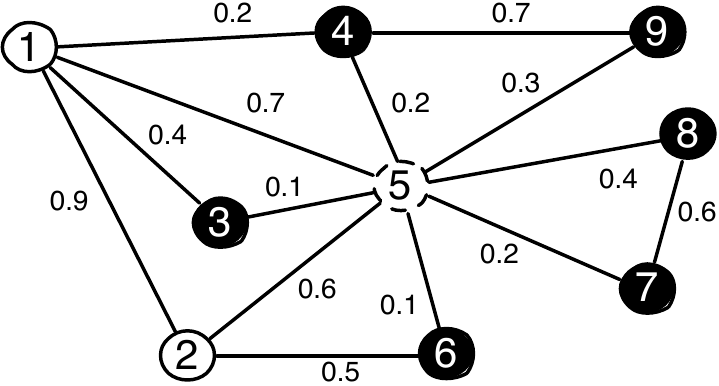}}
\caption{\small\bf Example of uncertain network. Nodes  $1$, $2$, are labeled {\em white}, and nodes $3$, $4$, $6$, $7$, $8$ and $9$ are labeled {\em black}, while the label of node $5$ is unknown and must be assigned by the algorithm. Edges between nodes exist with some probability.}
\label{fig:net2}
\vspace{0.1cm}
\end{center}
\end{figure}

Figure~\ref{fig:net2} shows an example.
Nodes  $1$, $2$, are labeled {\em white}, and nodes $3$, $4$, $6$, $7$, $8$
and $9$ are labeled {\em black}. The label of node $5$ is unknown and must
be assigned by the algorithm. We observe that ignoring the edges whose existence
probability is lower than $0.5$ is beneficial for the correct classification of node $5$.

Therefore, we use an iterative
augmentation process in order to reduce the impact of such edges,
by instead favoring the positive impact of high quality edges
in the collective classification process. The idea is to {\em
activate} only a subset of the edges for use on the modeling
process. In other words, edges which are not activated are not
used in the modeling.
We call this algorithm {\em uBayes+}.

We adopt a model inspired by automatic parameter selection in
machine learning. Note that, analogous to parameter selection, the
choice of a particular subset of high quality links, corresponds to
a configuration of the network, and we would like to determine an
optimal configuration for our approach.
In order to do this, we split the set of labeled nodes $T_0$ into two subsets: a {\em training set} denoted by $T_{train}$ and a {\em
hold out} set denoted by $T_{hold}$.
The ratio of the $T_0$ nodes that are assigned to the training set $T_{train}$ is
denoted by $\beta$, a user-defined parameter.

The purpose of the hold out set is to aid optimal configuration  selection by checking the
precise value of the parameters at which the training model provides
optimal accuracy over the set of nodes in $T_{hold}$. We use labels
of nodes in $T_{train}$ for the learning process, while using labels
of nodes in $T_{hold}$ as for the evaluation of accuracy at a
particular configuration of the network. (Note that a label is never
used for both the training and the hold out set, in order to avoid
overfitting.) The idea is to pick the ratio of active edges in such
a way so as to optimize the accuracy on the hold out set. This ensures
that an optimal fraction of the high quality edges are used for the
labeling process.

We start off considering a small fraction of the high probability
edges, iteratively expanding the subset of active edges by enabling
some of the inactive edges with the highest probabilities. The ratio
of active edges is denoted by the parameter $\theta$. Ideally, we
want to activate only the edges that contribute positively to the
classification of unlabeled nodes. Given a configuration of active
edges, we measure their goodness as the estimated accuracy on labels
of nodes in $T_{hold}$. The value of $\theta$ that leads to the
highest accuracy, denoted by $\theta^*$, is used as the ratio of
edges with the highest probability to activate on the uncertain
network $G$. The resulting network is then used as input for the
iterative probabilistic labeling algorithm ({\em uBayes}).

\hide{
{\bf CHARU NOTE: I note that $\alpha$ and
$\beta$ are not properly described below, and only pop out in
algorithm description. In the algorithm, they do not even seem to be
part of the  parameter set. We need to describe these parameters.
Also, are they optimized over using the hold out set? The
description here somehow does not seem to be complete. }
}

Despite optimizing accuracy by selecting the best ratio of edges to
be considered, the basic model described above is not very
efficient, because it requires multiple evaluations of the iterative
probabilistic labeling algorithm. In particular, it requires us to
vary the parameter $\theta$ and evaluate accuracy, in order to
determine $\theta^*$.

A more efficient technique for identifying $\theta^*$ can be
obtained by evaluating the accuracy for different values of $\theta$
on a sample of the uncertain network $G$ (rather than the full
network) as follows. We generate a new uncertain network $G' =
(N',A',P')$ by sampling  $\alpha \cdot |N|$ nodes from $G$ uniformly
at random, and retaining the edges from $A$ and probabilities from
$P$ referring to these sampled nodes. $\alpha$ is a user-defined
parameter that controls the ratio of nodes sampled from $G$ and it
implies the size of the sampled uncertain network $G'$. The initial
set of labeled nodes in the sampled uncertain network $G'$ is $T_0'
= T_0 \cap N'$. We split the set of nodes in $T_0'$ into two random
subsets, $T_{train}'$ and $T_{hold}'$, respectively. The number of
nodes in $T_{train}'$ is $\beta \cdot |T_0'|$.
We start off considering $\theta|A'|$ edges with the
highest probabilities, expanding iteratively the subset of active
edges at each iteration by increasing $\theta$. The goodness of
parameter $\theta$ is estimated as the accuracy of node labels in
$T_{hold}'$. Let $\theta^*$ be the value of $\theta$ leading to the
highest accuracy.  We activate $\theta^* |N|$ edges with highest
probability in $G$. The resulting network is then used as input for
the iterative probabilistic labeling (Algorithm~\ref{iterative}).
The overall algorithm is illustrated in Algorithm~\ref{iterative2}.

\begin{algorithm}
\begin{tabbing}
{\bf Algorithm} {\em uBayes+}(Graph: $G$\\
\ \ \ \ \ Uncertainty Prob.: $P$, Initial Labeling: $T_0$,\\
\ \ \ \ \ Sampled nodes ratio: $\alpha$, Train nodes ratio: $\beta$);\\ \\
{\bf begin}\\
\ \ \ \= $N'$ = Random sample of $\alpha \cdot |N|$ nodes from $N$;\\
\ \ \ \= $A'$ = Edges $(i,j)$ in $A$ with $i,j \in N'$;\\
\ \ \ \= $(T_{hold}',T_{train}') = split(T_0 \cap N', \beta)$;\\
\ \ \ \= $F = \theta \cdot |A'|$ edges in $A'$ with\\
\>\ \ \ greatest existence probability;\\
\> {\bf while} ($F \neq A'$) {\bf do}\\
\> {\bf begin}\\
\>\ \ \ \= Construct graph $G^F=(N',F)$;\\
\>\ \ \ \= $uBayes(G^F, P, T_{train})$;\\
\>\ \ \ \= Test accuracy using nodes in $T_{hold}$;\\
\>\ \ \ \= Expand edges in $F$ with top edges in $A'$;\\
\> {\bf end}\\
\>\ \ \ \= Construct graph $G^*=(N,F)$ with best\\
\>\ \ \ \= \ \  configuration (corresponding to $\theta^*$);\\
\>\ \ \ \= $uBayes(G^*, P, T_0)$;\\
{\bf end}
\end{tabbing}
\caption{Iterative Edge Augmentation for Uncertain Classification}
\label{iterative2}
\end{algorithm}

We note that the frequencies used to estimate conditional and prior probabilities
across the different configurations in Algorithm~\ref{iterative2}
can be efficiently maintained in an incremental fashion.

\subsection{Combining different classifiers}

In this section we propose a third algorithm, {\em uBayes+RN}. It
uses an ensemble methodology in order to further improve robustness
in scenarios, where some deterministic classifiers can provide good
results over {\em some} subsets of  nodes, but not over all the
nodes.  {\em uBayes+RN} is the linear combination of two
classifiers: the {\em uBayes+} algorithm and the Relational Neighbor
(RN) classifier \cite{macskassy2003simple}. The RN classifier is
defined as follows:

\begin{equation}
P_{RN}(L(r) = p) = \frac{1}{Z} \sum_{k : L(i_k) = p} p_{i_k r}
\end{equation}

where $p_{i_k r}$ is the probability and $Z= \sum_k p_{i_k r}$.
The \emph{uBayes+} and \emph{RN} algorithms are combined as follows:

\begin{eqnarray*}
P(L(r) = p) = \\
P(L(r)= p|L(i_1)=t_1 \ldots L(i_s)=t_s) \cdot \delta P_{RN}(L(r) = p)
\end{eqnarray*}

where $\delta$ controls the influence of the RN classifier during
the collective classification process. When $\delta = 0$ then
{\em uBayes+RN} degenerates to {\em uBayes+}, while when
$\delta = 1$ the two classifiers are weighted equally.
Note that this is a simple
linear combination. We used this combination, since it sometimes
provides greater robustness in the classification process.

%
%
%
%

\section{Complexity analysis}
\label{sec:complexity}

In this section, we discuss the complexity of the proposed algorithms.

We start with {\em uBayes}, which for the computation of the initial statistics requires $O(|N|+|A|)$ (label priors and conditional label probabilities).
Assuming that the cardinality of the set of immediate unlabeled neighbors of nodes in $T$ (remember that $T$ represents the set of currently labeled nodes) is at most $N_{max}$,
and that the number of neighbors for a particular node is at most $A_{max}$,
each iteration can be decomposed as follows.
The computation of new unlabeled nodes requires $O(N_{max} A_{max})$.
The computation of edge propagation probabilities requires $O(N_{max} A_{max})$.
The computation of node label probabilities requires $O(N_{max} A_{max})$.
Summing up, each iteration requires $O(N_{max} A_{max})$.
Assuming that all unlabeled nodes will be labeled in $K$ iterations,
the algorithm cost is $O(K |N| |A|)$, where $K << |N|$.
Space complexity is $O(|N| |A|)$.

For algorithm {\em uBayes+}, the computation of the uncertain network sample $G' = (N', A', P')$ requires $O(|N| + |A|)$.
Active edges are maintained using a priority list, whose initialization requires $O(|A|)$.
Each iteration of the iterative automatic parameter selection procedure
can be decomposed as follows.
Algorithm~\ref{iterative} (used by {\em uBayes+}) requires $O(|N'| |A'|)$.
Testing the classification accuracy requires $O(|N'|)$.
Expanding the set of active edges requires $O( |A'| log(|A'|))$.
Summing up, each iteration requires:

\begin{equation}
O(log(|A'|) |N'| |A'|).
\end{equation}

Finally, the last call to Algorithm~\ref{iterative} requires $O(K |N| |A|)$.
Assuming that the parameter selection procedure terminates after $K'$ iterations,
the algorithm cost is $O(|N| + |A| + K' (log(|A'|) |N'| |A'|))$.
Simplifying, the cost is $O(log(|A|) |N| |A|)$.
The space complexity is $O(|N| |A|)$.

Note that algorithms {\em uBayes+RN} and {\em uBayes+} have the same space and time complexity.

%
%
%
%

\section{Experimental Results}
\label{sec:results}

In this section, we evaluate the proposed techniques under different
settings, in terms of both accuracy and performance. 

We implemented
all techniques in C++ using the Standard Template Library (STL) and
Boost libraries, and ran the experiments on a Linux machine equipped
with an Intel Xeon 2.40GHz processor and 16GB of RAM. 

The reported
times do not include the initial loading time, which was constant
over all methods. The results were  obtained from $5$ independent
runs. For all experiments we report the averages and $95\%$
confidence intervals.

%
%
%
%

\subsection{Data Sets}
\label{sec:datasets} In our experiments, we used two  data sets
for which edge probabilities can be estimated, as described
below.

{\em DBLP:} The {\em DBLP data set} \cite{dblp12} is the most
comprehensive citation network of curated records of scientific
publications in computer science. In our experiments, we consider
the subset of publications  from $1980$ to $2010$. The data set
consists of $922,673$ nodes and $3,389,272$ edges. Nodes represent
authors and edges represent co-authorship relations. The edge
probability is an estimate of the probability that  two authors
co-authored a paper in a year selected randomly during their period
of activity.  For example, if a pair of authors  published papers in
ten different years and they both published papers for twenty years,
then their edge probability is $0.5$.
(We consider the union of their periods of activity.)
We  used  14 class labels,
that represent different research fields in computer science. The
corresponding labels and their frequencies are illustrated in Table~\ref{table:labels_dblp}.  The labels were generated by using  a set
of top conferences and journals in these areas, and the most frequent
label in the author's publications is used as the author's label. In our
data set, $16\%$ of the nodes are labeled. The rest were not labeled,
because the corresponding authors did not have publications in the
relevant conferences and journals.\\
\begin{table}[!htb]\small
\centering
\begin{tabular}{|c|c|c|} \hline
Id & Name & Prior probability \\ \hline \hline
$C_1$ & Verification \& Testing & $0.06758$\\
$C_2$ & Computer Graphics & $ 0.01974$\\
$C_3$ & Computer Vision & $0.04144$\\
$C_4$ & Networking & $0.1301$\\
$C_5$ & Data Mining & $0.09498$\\
$C_6$ & Operating systems & $0.06058$\\
$C_7$ & Computer Human Interaction & $0.05361$\\
$C_8$ & Software Engineering & $0.01935$\\
$C_9$ & Machine Learning & $0.1543$\\
$C_{10}$ & Bioinformatics & $0.1936$\\
$C_{11}$ & Computing Theory & $0.04008$\\
$C_{12}$ & Information Security & $0.05364$\\
$C_{13}$ & Information Retrieval & $0.044$\\
$C_{14}$ & Computational Linguistics & $0.02711$\\
\hline
\end{tabular}
\caption{Node labels and label priors of the {\em DBLP} dataset.}
\label{table:labels_dblp} 
\end{table}

\begin{table}[!htb]\small
\centering
\begin{tabular}{|c|c|c|} \hline
Id & Name & Prior probability \\ \hline \hline
$C_1$ & Verification \& Testing & $0.06758$\\
$C_2$ & Computer Graphics & $ 0.01974$\\
$C_3$ & Computer Vision & $0.04144$\\
$C_4$ & Networking & $0.1301$\\
$C_5$ & Data Mining & $0.09498$\\
$C_6$ & Operating systems & $0.06058$\\
$C_7$ & Computer Human Interaction & $0.05361$\\
$C_8$ & Software Engineering & $0.01935$\\
$C_9$ & Machine Learning & $0.1543$\\
$C_{10}$ & Bioinformatics & $0.1936$\\
$C_{11}$ & Computing Theory & $0.04008$\\
$C_{12}$ & Information Security & $0.05364$\\
$C_{13}$ & Information Retrieval & $0.044$\\
$C_{14}$ & Computational Linguistics & $0.02711$\\
\hline
\end{tabular}
\caption{Class labels and corresponding conference keywords}
\label{table:keywords_dblp} 
\end{table}
{\em US Patent Data Set:} The US Patent data set \cite{hall2001nber}
is a citation network of US utility patents. In our experiments, we
consider patents issued from $1970$ to $1990$. The network contained
$108,658$ nodes and $1,059,822$ edges. A node represents a patent
assignee and there is an edge between two assignees if there is at
least a patent from one assignee citing a patent from the other
assignee. The edge probability is an estimate of the probability
that one of the two assignee cites the other assignee.  For example,
assignee $A$ cites $20$ patents of which $5$ are assigned to
assignee $B$, then their edge probability is $0.25$. A category is
assigned to each patent. The most frequent category in the
assignee's  patents is used as assignee label.
Table~\ref{table:labels_patent} reports the label class names and
their frequencies.   These labels cover $66\%$ of nodes. We used
class label as ground truth. Although the raw input data sets are
curated manually, class labels are derived algorithmically and may
be noisy. For example, if an assignee holds only two patents
belonging to different categories, we pick  one of these two
categories randomly as the assignee label. In other words, we do not
model our confidence in the derived class labels. Results show that
the proposed algorithms is  robust to this lack of information.

\begin{table}[!htb]\small
\centering
\begin{tabular}{|c|c|c|} \hline
Id & Name & Prior probability \\ \hline \hline
$C_1$ & Chemical & $0.2077$ \\
$C_2$ & Computers \& communications & $0.07945$ \\
$C_3$ & Drugs \& Medical & $0.0859$ \\
$C_4$ & Electrical \& Electronic & $0.19292$ \\
$C_5$ & Mechanical & $0.434$ \\
\hline
\end{tabular}
\caption{Node labels and label priors of the {\em Patent} data set.}
\label{table:labels_patent} 
\end{table}

%
%
%
%

\subsection{Perturbation}
\label{subsec:perturbation}

We also used  perturbed data sets to stress-test the methods. The
advantage of such data is the ability to test the effectiveness with
varying uncertainty level, and other sensitivity parameters. This
provides a better idea of the inherent variations of the
performance. Perturbed data sets are generated by either adding
noisy edges or by removing existing edges to and from the real data
sets. Noisy edges are new edges with low probability. The edge
probability is sampled from a normal distribution $N(0,\sigma)$ in
the interval $(0,1]$. The parameter $\sigma$ controls the
probability standard deviation. As it gets larger the average edge
probability increases, eventually interfering with edges in the real
data sets. The parameter  $\phi$ controls the ratio of noisy edges.
Given the edge set $A$ of a real data set, the number of added noisy
edges is $\phi \cdot |A|$.

The existing edges to be removed are selected by sampling the edge set $A$ uniformly at random.
Existing edges are removed {\em after} adding noisy edges.
The parameter $\Phi$ controls the ratio of edges to be removed.
Given the edge set $A$ of a perturbed data set, the number of retained
edges is $(1-\Phi) |A|$. The selection criterion is also known as
{\em probability sampling}.

The existing labeled nodes to be unlabeled are selected by sampling
the node set $N$ randomly. The parameter $\Gamma$ controls the ratio
of labeled nodes, whose label is to be removed. Given the node set
$N$ of a real data set, the number of labeled nodes whose label is
removed is $(1-\Gamma) |N|$.

Unless otherwise  specified, we used  the following default
perturbation parameters. The ratio of noisy edges ($\phi$) is $3$
and the standard deviation of noisy edges ($\sigma$) is $0.25$. By
default, we do not remove any edges or labels. Thus, $\Phi$ equals
zero, and the ratio of known labels for the data sets are those
reported in Section~\ref{sec:datasets}.

\subsection{Evaluation Methodology}

The accuracy is assessed by using repeated random sub-sampling
validation. We randomly partition the nodes into training and
validation subsets which are denoted by  $N_T$ and $N_V$
respectively. We use $2/3$ of the labeled nodes for training, and
the remaining $1/3$ for validation.
Even if $2/3$ may appear as a large fraction, note that
it refers to the labeled nodes in the ground truth (that is rather limited).

For each method, we compute the confusion matrix $M$ on the $N_V$
set, where $M_{ij}$ is the count of nodes labeled as $i$ in the
ground-truth that are
 labeled as $j$. Accuracy is defined as the ratio of true positives for
all class labels:
\begin{equation}
Accuracy =  \frac{1}{|N_V|} \sum M_{ii}
\end{equation}

If all nodes are labeled correctly, $M$ is a diagonal matrix.
The experiment is repeated several times to get statistically
significant results.

In all experiments we use the following parameters for the {\em uBayes+} algorithm.
The ratio nodes of the sampled uncertain network ($\alpha$) is $0.2$.
Among the sampled nodes, the ratio nodes used for training ($\beta$) is $0.1$.
In order to identify $\theta^*$, the algorithm varies $\theta$ between $0.05$ and $1$ in $20$ steps.
These values were determined experimentally, and are the same for both datasets (the performance of the algorithms remains stable for small variations of these parameters).

We compared our techniques to two algorithms, which are the  {\em
wvRN} \cite{macskassy2003simple} and {\em Sampling} methods. 
Since these algorithms trade accuracy for running time, we limited the
running time of these two algorithms to the time spent by the {\em
uBayes} method.  The {\em wvRN} method estimates the probability of
node $i$ to have label $j$ as the weighted sum of  class membership
probabilities of neighboring nodes for label $j$.   Thus, it works
with a weighted deterministic representation of the network, where
the edge probabilities are used as weights. Relaxation labeling is
then used for inference. We additionally consider a version of the
{\em wvRN} algorithm, {\em wvRN-20}, that is not time-bounded, but
is bound to terminate after $20$ iterations in the label relaxation
procedure. As we discuss later, the accuracy of {\em wvRN} converges
quickly and does not improve further after $20$ iterations for both
data sets.

The sampling algorithm samples networks in order to create
deterministic representations. For each sampled instantiation, the
{\em RN} algorithm \cite{macskassy2003simple} is used. Note that
links in sampled instantiations either exist or do not exist, and
link weights are set to $1$. This algorithm  estimates class
membership probabilities by voting on the different labelings over
different instantiations of the network. The class with the largest
vote is reported as the relevant label.

\subsection{Classification Quality Results}

\begin{figure*}[tbh]
\centering
\vspace{0.1cm}
\begin{minipage}{0.32\linewidth}
\begin{tabular}{c}
\includegraphics[angle=0,scale=0.22]{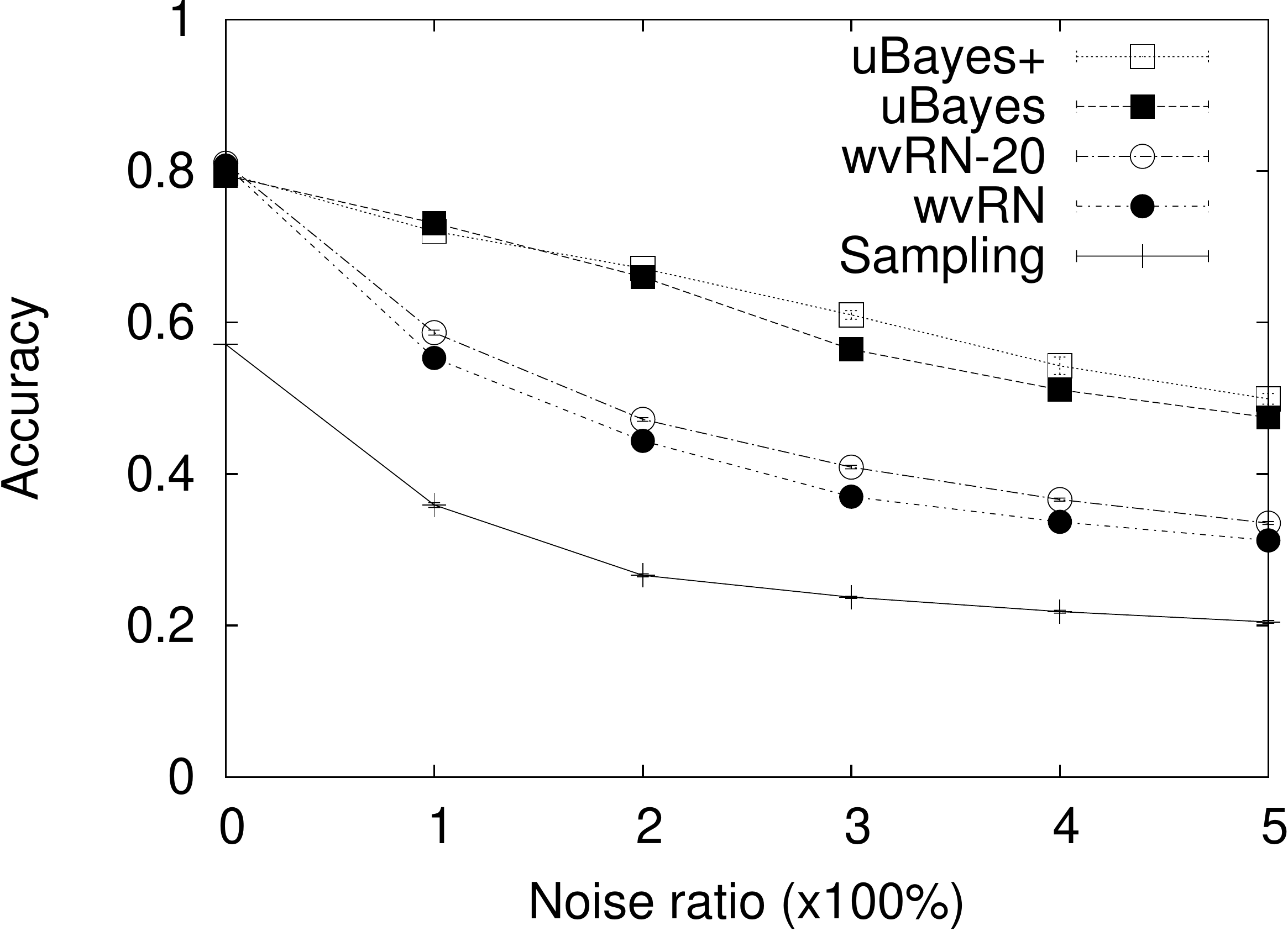} \\
\\
\\
(a) DBLP\\
\includegraphics[angle=0,scale=0.22]{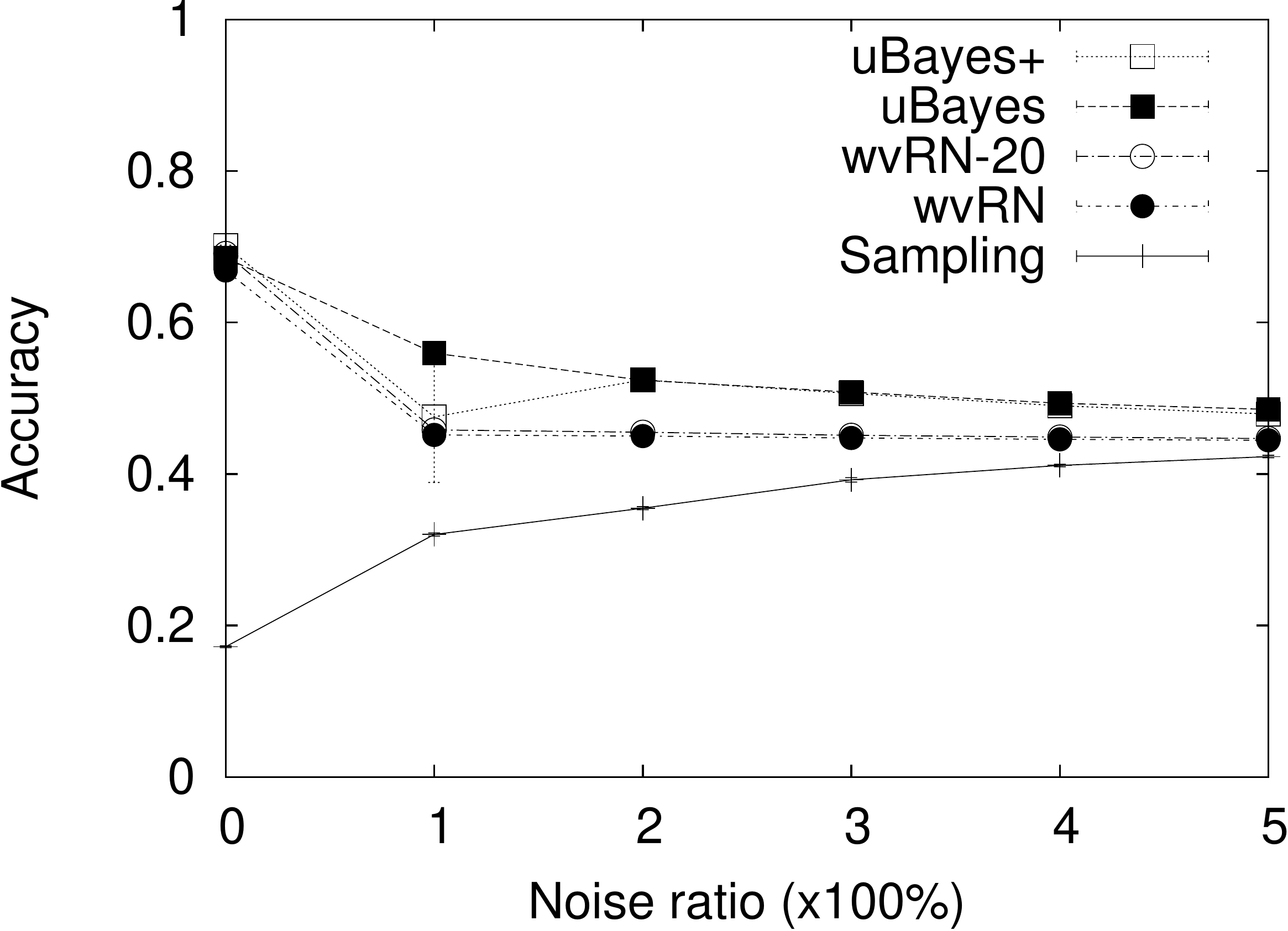} \\
\\
\\
(b) Patent\\
\end{tabular}
\caption{\small\bf Accuracy  with varying ratio of noisy edges for
algorithms {\em uBayes}, {\em uBayes+}, {\em wvRN}, {\em wvRN-20}
and {\em Sampling}. } \label{fig:gammaacc}
\end{minipage}
\hspace{0.005\linewidth}
\begin{minipage}{0.32\linewidth}
\begin{tabular}{c}
\includegraphics[angle=0,scale=0.22]{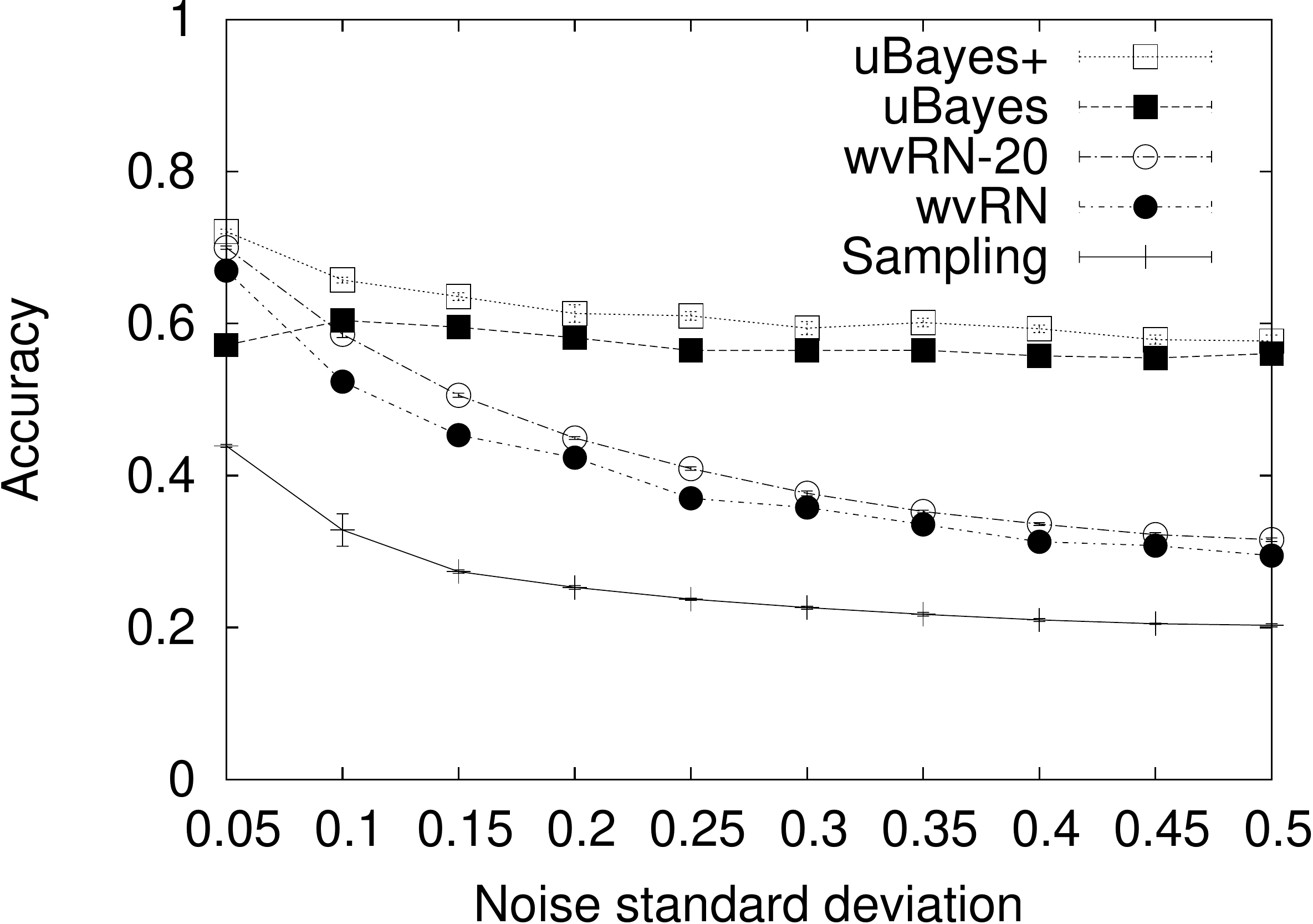} \\
\\
\\
(a) DBLP\\
\includegraphics[angle=0,scale=0.22]{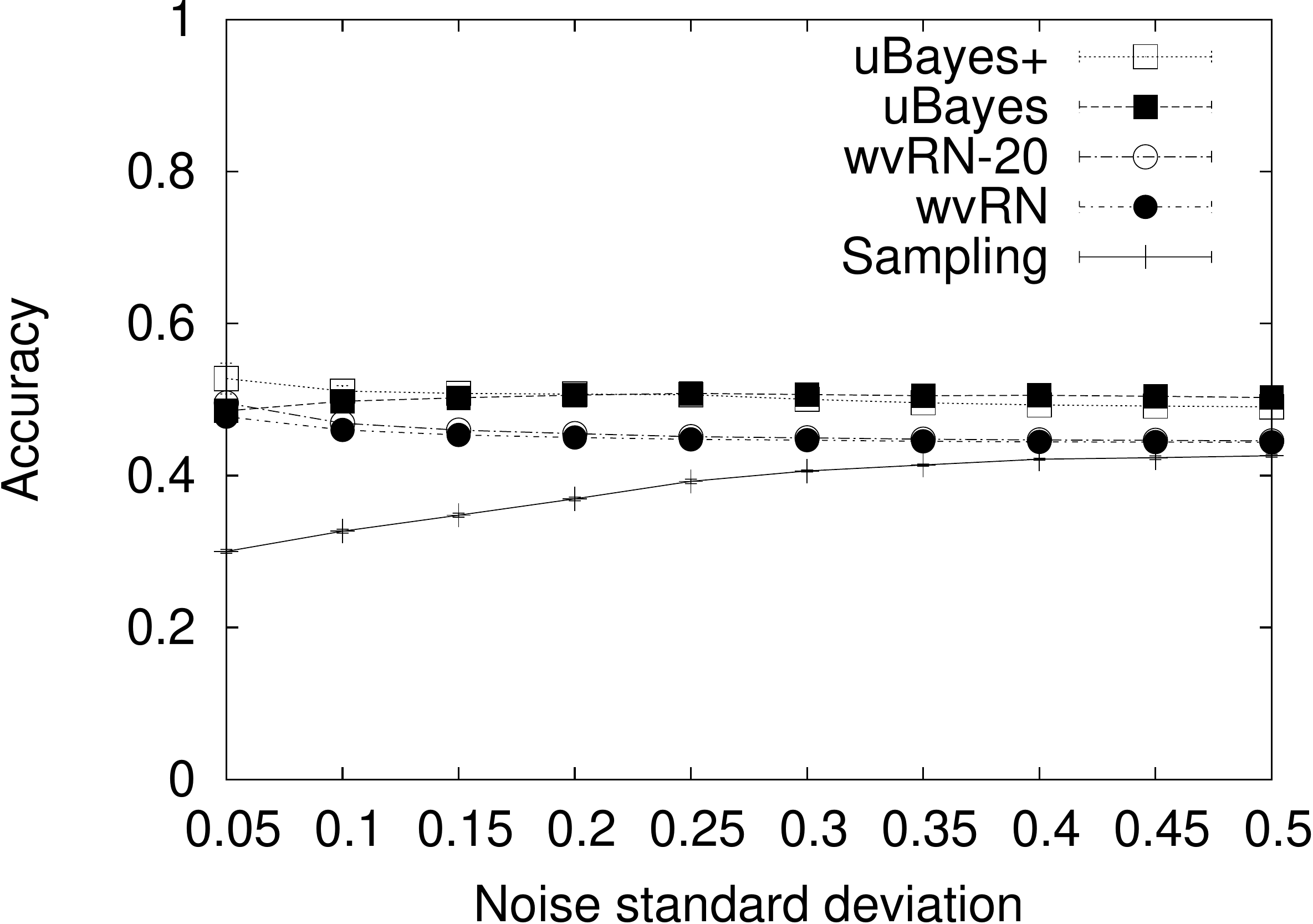} \\
\\
\\
(b) Patent\\
\end{tabular}
\caption{\small\bf Accuracy with  varying standard deviation of
probability of noisy edges for {\em uBayes}, {\em uBayes+}, {\em
wvRN}, {\em wvRN-20} and {\em Sampling} algorithms.}
\label{fig:thetaacc}
\end{minipage}
\hspace{0.005\linewidth}
\begin{minipage}{0.32\linewidth}
\begin{tabular}{c}
\includegraphics[angle=0,scale=0.22]{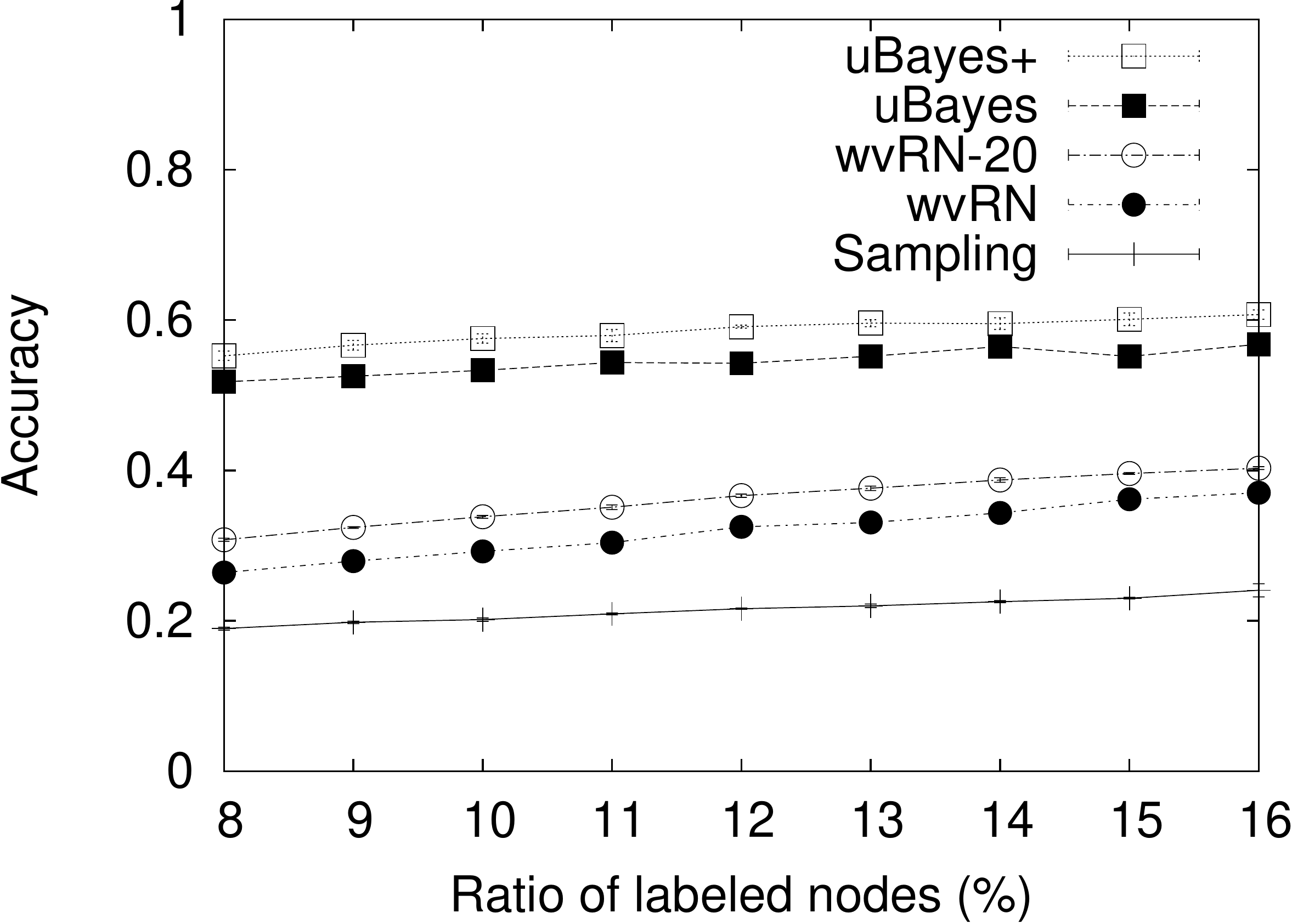} \\
\\
\\
(a) DBLP\\
\includegraphics[angle=0,scale=0.22]{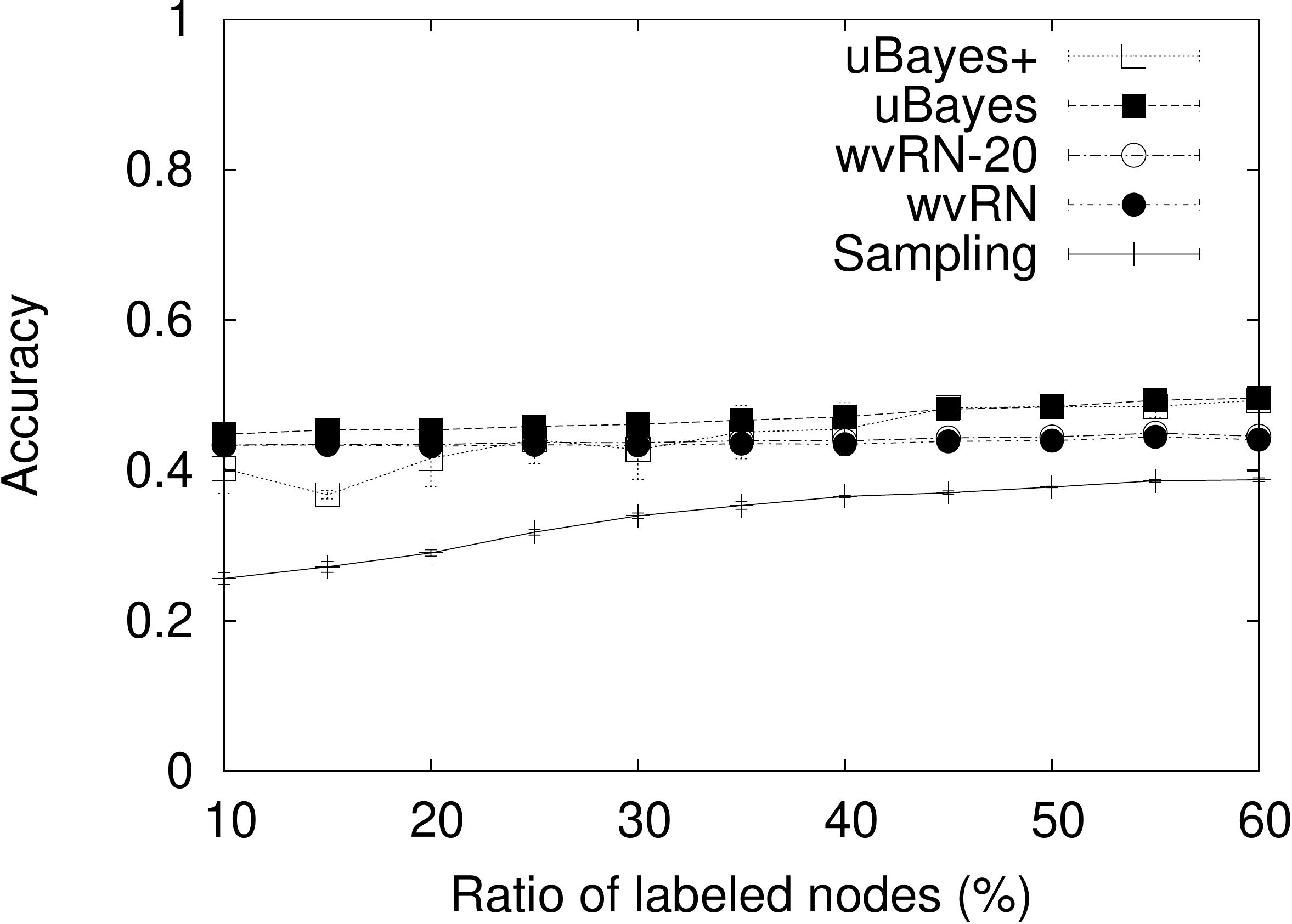} \\
\\
\\
(b) Patent\\
\end{tabular}
\caption{\small\bf Accuracy with  varying ratio of labeled nodes for
{\em uBayes}, {\em uBayes+}, {\em wvRN}, {\em wvRN-20} and {\em
Sampling} algorithms.} \label{fig:labeledacc}
\end{minipage}
\vspace{0.1cm}
\end{figure*}

In this section, we report our results on accuracy under a variety
of settings using both real and perturbed data sets.  The first
experiment shows the accuracy by varying the ratio of noisy edges
($\phi$) for the algorithms {\em uBayes}, {\em uBayes+}, {\em wvRN},
{\em wvRN-20} and {\em Sampling}. The results for the {\em DBLP} and
{\em Patent} data sets are reported in Figures~\ref{fig:gammaacc}(a)
and \ref{fig:gammaacc}(b), respectively.  The {\em Sampling}
algorithm is the worst performer on both data sets, followed by {\em
wvRN} and {\em wvRN-20}. On the {\em DBLP} data set, the accuracy of
{\em wvRN-20} is slightly higher than that of  {\em wvRN}.  The {\em
uBayes} and {\em uBayes+} algorithms are the best performers, with
{\em uBayes+} achieving higher accuracy on the {\em DBLP} dataset
when the ratio of noisy edges is above $200\%$. We observe that
there is nearly no difference among the {\em uBayes}, {\em uBayes+},
{\em wvRN} and {\em wvRN-20} algorithms on both datasets when $\phi
= 0$, while the percentage improvement in accuracy from {\em
wvRN-20} to {\em uBayes+} when $\phi = 5$ ($500\%$) is up to $49\%$
for {\em DBLP} and $7\%$ for {\em Patent}. It is worth noting that,
as the ratio of noisy edges increases, the accuracy for {\em
Sampling} increases in the Patent data set. This is due to the high
probability of label $C_5$ ($0.434$), as reported in
Table~\ref{table:labels_patent}, which eventually dominates the
process.

In the next experiment, we varied the standard deviation of the
probability of the noisy edges ($\sigma$) for algorithms {\em
uBayes}, {\em uBayes+}, {\em wvRN}, {\em wvRN-20} and {\em
Sampling}. The results for the {\em  DBLP} and the {\em Patent} data
sets are reported in Figures~\ref{fig:thetaacc}(a) and
\ref{fig:thetaacc}(b), respectively.  The {\em Sampling} algorithm
again does not perform well, followed by the {\em wvRN} and {\em
wvRN-20} algorithms. The  {\em uBayes+}  algorithm is consistently
the best performer on the {\em DBLP} dataset, while there is nearly
no difference between the {\em uBayes+} and {\em uBayes}  algorithms
on the {\em Patent} data set. The higher accuracy of {\em uBayes}
and {\em uBayes+} is explained by their ability to better capture
correlations between different class labels, a useful feature when
processing noisy data sets. The better performance of {\em uBayes+}
is due to its ability to ignore noisy labels that contribute
negatively to the overall classification process. {\em uBayes+} is
more accurate than {\em wvRN-20} with a percentage improvement up to
$83\%$ in the {\em DBLP} data set and $10\%$ in the {\em Patent}
data set, which represents a significant advantage.

In the following experiment, we evaluate the accuracy when varying
the ratio of labeled nodes ($\Gamma$) for algorithms {\em uBayes},
{\em uBayes+}, {\em wvRN}, {\em wvRN-20} and {\em Sampling}.
(Default perturbation parameters are considered for the retained edges.)
The results for the  {\em DBLP} and {\em Patent} data sets are reported
in Figures~\ref{fig:labeledacc}(a) and \ref{fig:labeledacc}(b)
respectively. In the {\em DBLP} dataset,  the {\em wvRN} algorithm
performs better than {\em wvRN-20}, while there is virtually no
difference on the {\em Patent} dataset.  The {\em uBayes+} algorithm
is consistently the best performer on the {\em DBLP} dataset, while
it performs slightly worse than {\em uBayes} on the {\em Patent}
data set when $\Gamma$ is below $0.2$ ($20\%$). We observe that the
percentage improvement of {\em uBayes+} over {\em wvRN-20} is $50\%$
on the {\em DBLP} dataset and $11\%$ on the {\em Patent} dataset.
The {\em Sampling} algorithm exhibits the lowest accuracy.

\begin{figure*}[tbh]
\centering
\vspace{0.1cm}
\begin{minipage}{0.32\linewidth}
\begin{tabular}{c}
\includegraphics[angle=0,scale=0.22]{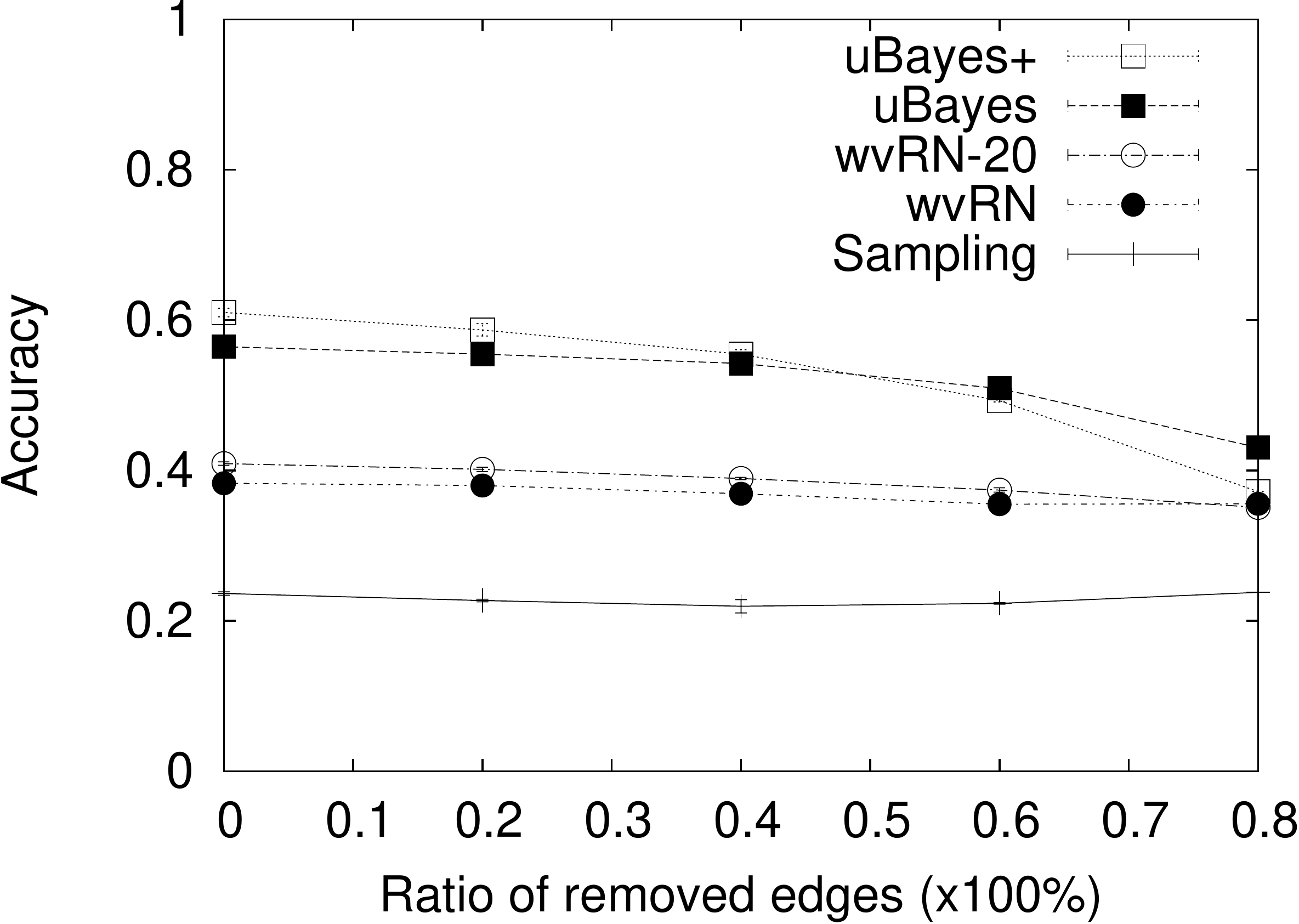} \\
\\
\\
(a) DBLP\\
\includegraphics[angle=0,scale=0.22]{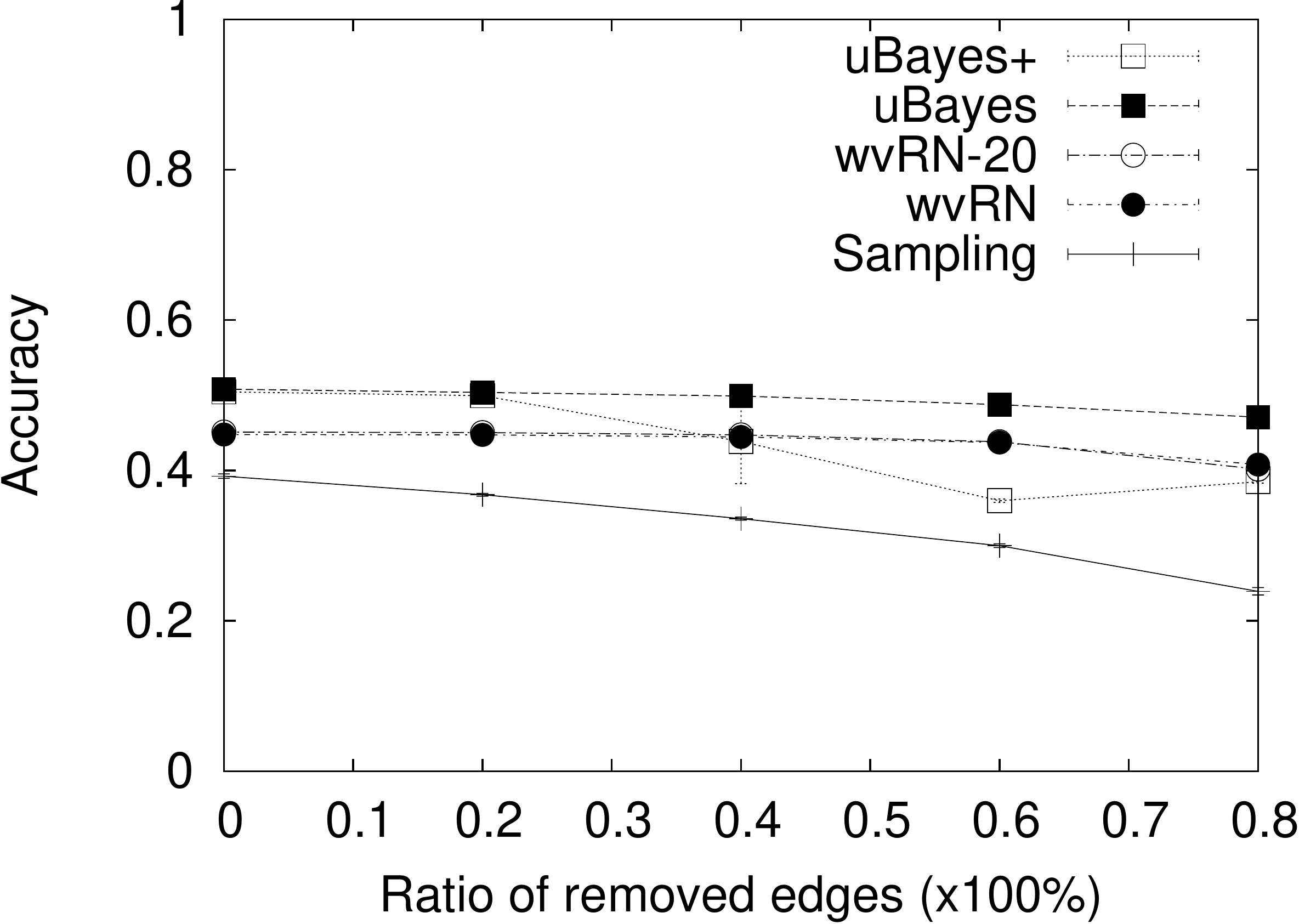} \\
\\
\\
(b) Patent\\
\end{tabular}
\caption{\small\bf Accuracy with varying  ratio of retained edges
for \emph{uBayes}, \emph{uBayes+}, {\em wvRN}, {\em wvRN-20} and
{\em Sampling} algorithms.} \label{fig:ratioedges}
\end{minipage}
\hspace{0.005\linewidth}
\begin{minipage}{0.32\linewidth}
\begin{tabular}{c}
\includegraphics[angle=0,scale=0.22]{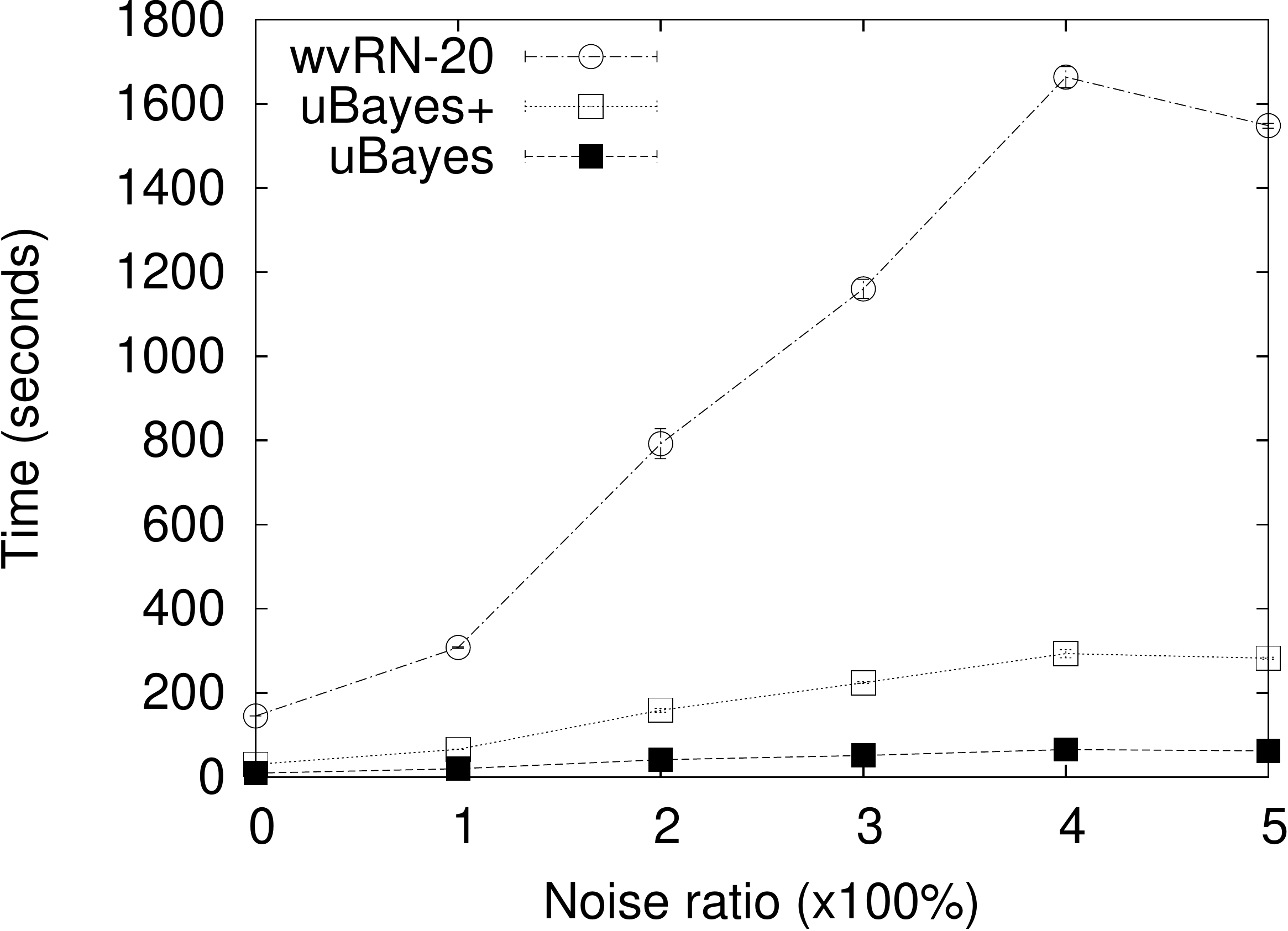} \\
\\
\\
(a) DBLP\\
\includegraphics[angle=0,scale=0.22]{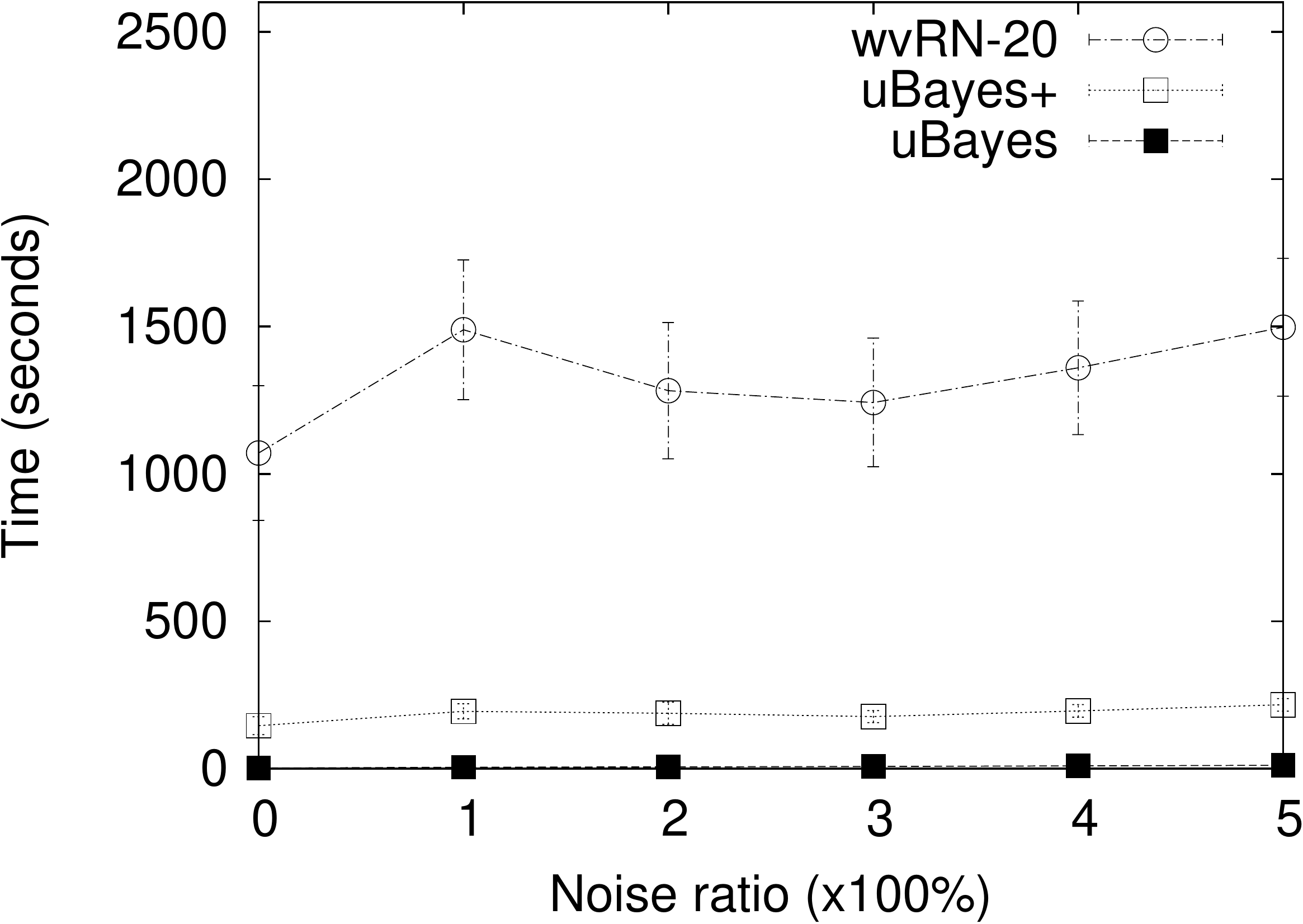} \\
\\
\\
(b) Patent\\
\end{tabular}
\caption{\small\bf Time performance with  varying ratio of noisy
edges for {\em uBayes}, {\em uBayes+} and {\em wvRN-20} algorithms.}
\label{fig:time_gamma}
\end{minipage}
\hspace{0.005\linewidth}
\begin{minipage}{0.32\linewidth}
\begin{tabular}{c}
\includegraphics[angle=0,scale=0.22]{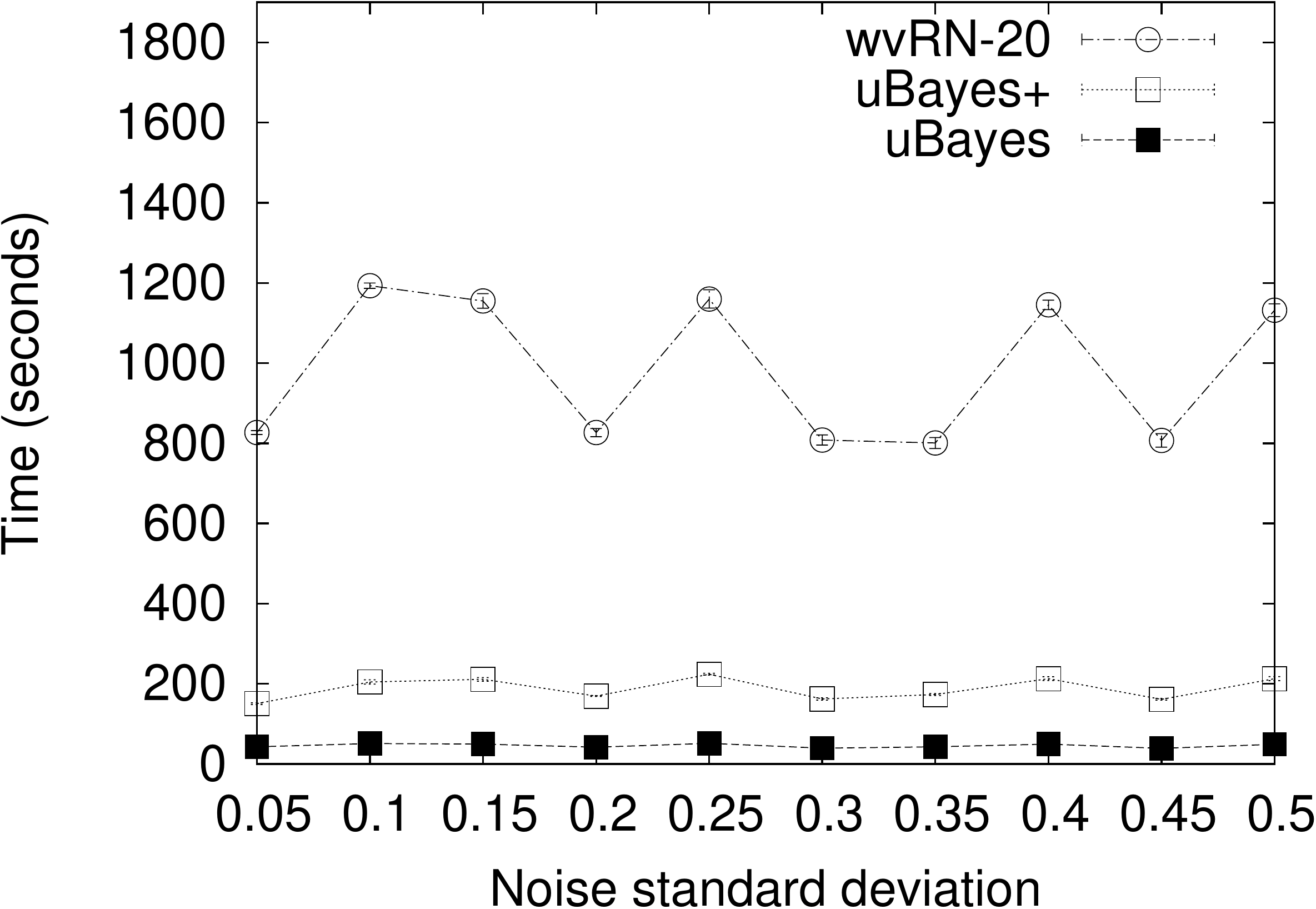} \\
\\
\\
(a) DBLP\\
\includegraphics[angle=0,scale=0.22]{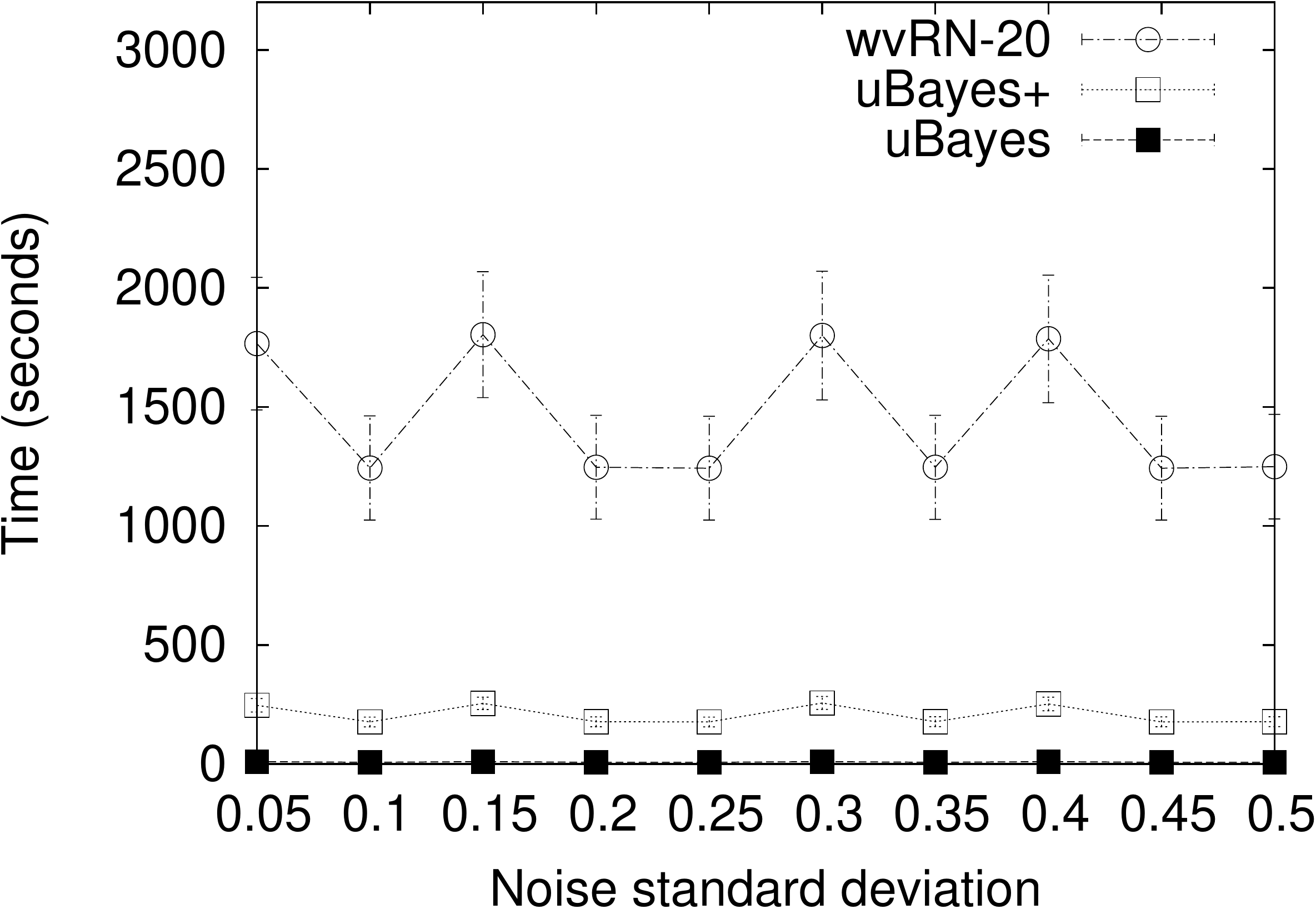} \\
\\
\\
(b) Patent\\
\end{tabular}
\caption{\small\bf Time performance with varying standard deviation
of probability of noisy edges for \emph{Bayes}, \emph{Bayes+} and
{\em wvRN-20} algorithms.} \label{fig:time_theta}
\end{minipage}
\vspace{0.1cm}
\end{figure*}

We now stress-test the proposed techniques by randomly removing a
percentage of edges ($\Phi$), as detailed in
Section~\ref{subsec:perturbation}. The results for the {\em DBLP}
and the {\em Patent} data sets are reported  in
Figures~\ref{fig:ratioedges}(a) and \ref{fig:ratioedges}(b),
respectively. {\em Sampling}  consistently performs at the lower
range, followed by {\em wvRN} and {\em wvRN-20}. {\em uBayes} and
{\em uBayes+} perform consistently better on the {\em DBLP} dataset,
with {\em uBayes+} performing poorly when the ratio of retained
edges is below $60\%$.  In  this case, the resulting network is less
connected, and the uncertain network sample used for the automatic
parameter tuning becomes less robust to noisy conditions. In the
{\em DBLP} dataset, the percentage improvement of {\em uBayes+} over
{\em wvRN-20} is up to $49\%$.

In Table~\ref{table:cm_dblp}, we report the confusion matrices for
the {\em DBLP } and {\em Patent} data sets for the {\em uBayes+}
algorithm.  The confusion matrix provides some interesting insights,
especially for cases where nodes were misclassified. Cell $i,j$
reports the number of nodes with ground truth label $C_i$ classified
with label $j$. We observe that $C_4$, $C_9$ and $C_{10}$ labels
(networking, machine-learning and bioinformatics) in the {\em DBLP}
data set lead to many misclassifications. This can be explained by
the fact that these are the most frequent labels in the network
(refer to Table~\ref{table:labels_dblp}), and therefore have a
higher probability of being selected. We also observe that
misclassifications convey interesting and useful information. For
example, excluding the $C_4$, $C_9$ and $C_{10}$ classes, most of
the misclassifications for class ``Data Mining" are due to the
``Information Retrieval" class, and vice versa. This points to the fact that the
two communities are related to each other. Similar observations can
be made on the {\em Patent} data set. For example, the ``Chemical"
and ``Drugs \& Medical" classes overlap, and show corresponding
behavior in the confusion matrices.

\begin{table*}[!htb]\small
\vspace{0.5cm}
\centering
\begin{tabular}{|c|c|c|c|c|c|c|c|c|c|c|c|c|c|c|} \hline
&$C_{1}$&$C_{2}$&$C_{3}$&$C_{4}$&$C_{5}$&$C_{6}$&$C_{7}$&$C_{8}$&$C_{9}$&$C_{10}$&$C_{11}$&$C_{12}$&$C_{13}$&$C_{14}$\\ \hline \hline
$C_{1}$&$\mathbf{1943}$&$7$&$9$&$116$&$19$&$58$&$11$&$38$&$364$&$595$&$32$&$35$&$4$&$7$\\
$C_{2}$&$3$&$\mathbf{620}$&$30$&$21$&$12$&$7$&$37$&$3$&$125$&$131$&$6$&$4$&$3$&$1$\\
$C_{3}$&$15$&$49$&$\mathbf{1088}$&$51$&$25$&$18$&$13$&$3$&$489$&$261$&$11$&$8$&$11$&$3$\\
$C_{4}$&$55$&$21$&$22$&$\mathbf{3907}$&$70$&$161$&$38$&$19$&$915$&$1076$&$69$&$87$&$23$&$8$\\
$C_{5}$&$12$&$17$&$26$&$182$&$\mathbf{2237}$&$92$&$49$&$45$&$960$&$791$&$60$&$79$&$126$&$37$\\
$C_{6}$&$60$&$26$&$12$&$311$&$86$&$\mathbf{1482}$&$26$&$39$&$423$&$464$&$105$&$80$&$35$&$7$\\
$C_{7}$&$7$&$36$&$13$&$43$&$14$&$12$&$\mathbf{1396}$&$20$&$463$&$584$&$4$&$15$&$39$&$15$\\
$C_{8}$&$7$&$3$&$0$&$18$&$10$&$15$&$18$&$\mathbf{619}$&$113$&$128$&$4$&$8$&$7$&$4$\\
$C_{9}$&$37$&$46$&$271$&$279$&$152$&$47$&$72$&$23$&$\mathbf{5422}$&$1287$&$74$&$85$&$110$&$104$\\
$C_{10}$&$42$&$32$&$81$&$274$&$127$&$80$&$45$&$26$&$692$&$\mathbf{8410}$&$79$&$58$&$44$&$45$\\
$C_{11}$&$28$&$10$&$7$&$135$&$22$&$168$&$9$&$13$&$258$&$321$&$\mathbf{950}$&$46$&$5$&$3$\\
$C_{12}$&$27$&$13$&$18$&$236$&$62$&$70$&$33$&$26$&$428$&$483$&$35$&$\mathbf{1230}$&$20$&$7$\\
$C_{13}$&$7$&$9$&$14$&$73$&$94$&$22$&$67$&$29$&$452$&$415$&$15$&$18$&$\mathbf{873}$&$82$\\
$C_{14}$&$5$&$5$&$5$&$40$&$20$&$8$&$14$&$2$&$268$&$226$&$12$&$11$&$34$&$\mathbf{756}$\\
\hline
\end{tabular}
\caption{Confusion matrix for {\em DBLP} dataset. True positives are
indicated in bold.} \label{table:cm_dblp}
\vspace{0.1cm}
\end{table*}

\begin{table}[!htb]\small
\centering
\vspace{0.1cm}
\begin{tabular}{|c|c|c|c|c|c|} \hline
&$C_{1}$&$C_{2}$&$C_{3}$&$C_{4}$&$C_{5}$\\ \hline \hline
$C_{1}$&$\mathbf{939}$&$5$&$34$&$79$&$3861$\\
$C_{2}$&$21$&$\mathbf{328}$&$4$&$98$&$1421$\\
$C_{3}$&$181$&$0$&$\mathbf{279}$&$26$&$1567$\\
$C_{4}$&$96$&$108$&$26$&$\mathbf{574}$&$3798$\\
$C_{5}$&$292$&$63$&$26$&$133$&$\mathbf{10020}$\\
\hline
\end{tabular}
\caption{Confusion matrix for {\em Patent} data set. True positives
are indicated in bold.} \label{table:cm_patent}
\vspace{0.1cm}
\end{table}

Finally, we report  the accuracy of the {\em uBayes+RN} algorithm
when varying the parameter $\delta$ between $0$ and $1$. Recall that
$\delta$ controls the influence of the {\em RN} classifier on the
overall classification process. In our experiments with both data
sets, the accuracy of {\em uBayes+RN} was always slightly better
than {\em uBayes+}, but never more than $5\%$.
We observed nearly no difference among the different $\delta$ configurations.
In the interest of space, we omit the detailed results.

We next provide some  real examples of labeling results obtained
with {\em uBayes+} on the Patent dataset. The ``{\em Atari Inc.}"
and ``{\em Sega Enterprises, Ltd}" companies, which belong to the
hall of fame of the video game industry, were not assigned to any
category. Our algorithm correctly classified them as ``Computers \&
Communications". Similarly, the companies ``{\em North American
Biologicals, Inc}" and  ``{\em Bio-Chem Valve, Inc}" were correctly
labeled as ``Drugs \& Medical", since they are both involved in drug
development and pharmaceutical research. Interestingly, ``{\em
Starbucks Corporation}" was labeled as ``Chemical". Taking a close
look at their patents, it turns out that a large fraction of them
describe techniques for enhancing flavors and aromas that involve
chemical procedures.
Evidently, having labels for all the nodes in the graph allows for
improved query answering and data analysis in general.

%
%
%
%

\subsection{Efficiency Results}

In this section, we assess running time efficiency on a variety of
settings using both real and perturbed data sets.
Figures~\ref{fig:time_gamma}(a) and \ref{fig:time_gamma}(b) show the
CPU time required by the algorithms when varying the ratio of noisy
edges, for the {\em DBLP} and {\em Patent} data sets, respectively.
Note that {\em Sampling} has the same time performance as {\em
uBayes}. The {\em uBayes+} algorithm is nearly three times slower
than {\em uBayes}. This is due to the automatic parameter tuning
approach employed by the  {\em uBayes+} algorithm We observe that
the performance of {\em wvRN-20} almost always {\em considerably}
worse than both {\em uBayes} and {\em uBayes+}. The same observation
is true when we vary the standard deviation of the probability of
the noisy edges (see Figures~\ref{fig:time_theta}(a) and
\ref{fig:time_theta}(b)). Note that the inference in the {\em wvRN}
algorithm is based on labeling relaxation, whose complexity is
proportional to the size of the network and remains constant across
iterations. On the contrary, the iterative labeling that {\em
uBayes} and {\em uBayes+} use for their inference model becomes
faster with each successive iteration, since it needs to visit a
smaller part of the network. As the results show, the standard
deviation does not affect the time performance of the algorithms.
These experiments demonstrate that the two proposed algorithms
effectively combine low running times with high accuracy and
robustness levels.

\begin{figure}[htb!]
\center
\vspace{0.1cm}
\includegraphics[angle=0,scale=0.3]{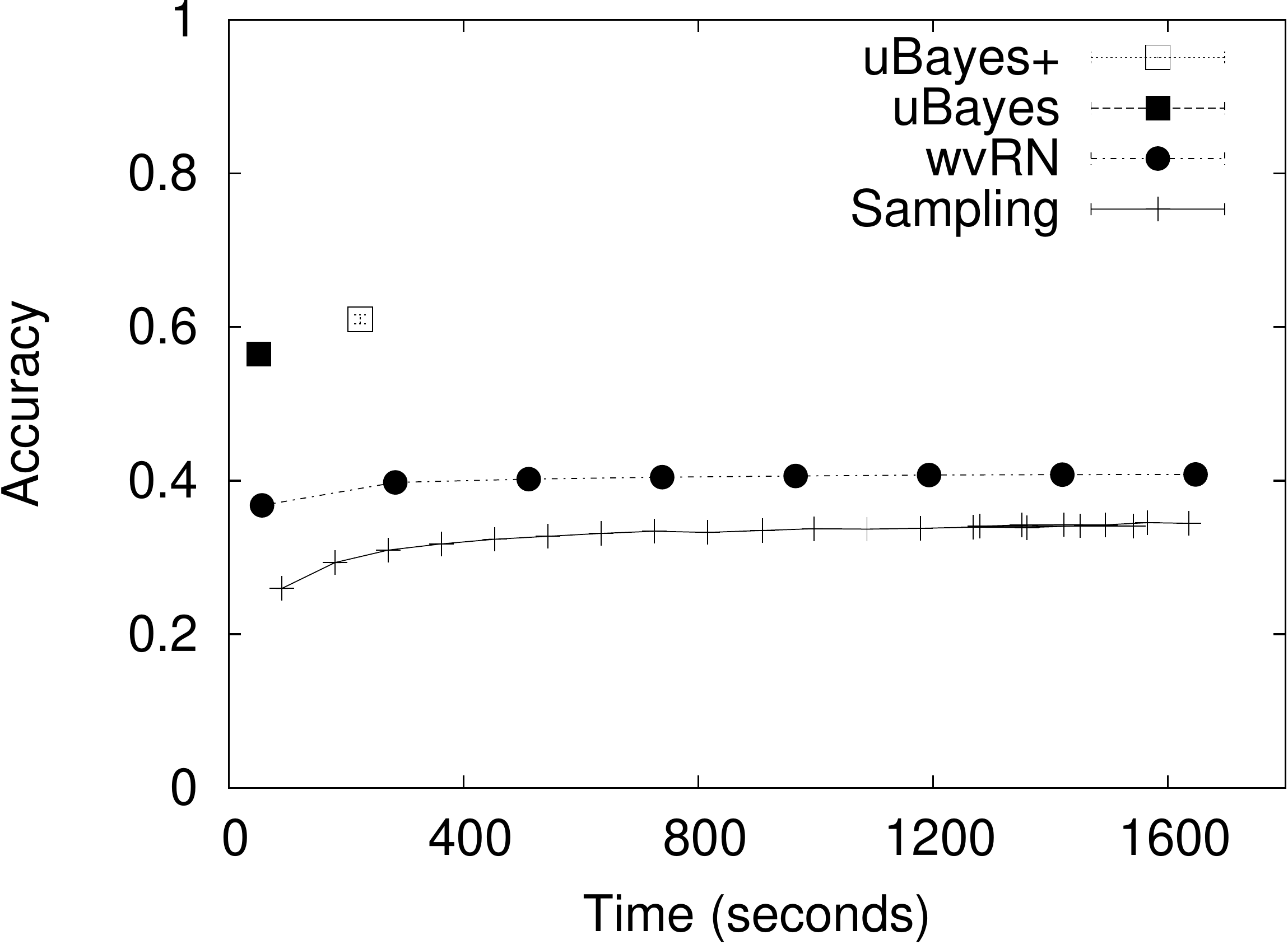}
\caption{\small\bf Accuracy with varying execution times for {\em
uBayes}, {\em uBayes+}, {\em wvRN} and {\em Sampling} algorithms for
the {\em DBLP} data set.} \label{fig:dblp_timeacc}
\vspace{0.1cm}
\end{figure}
\begin{figure}[htb!]
\center
\vspace{0.1cm}
\includegraphics[angle=0,scale=0.3]{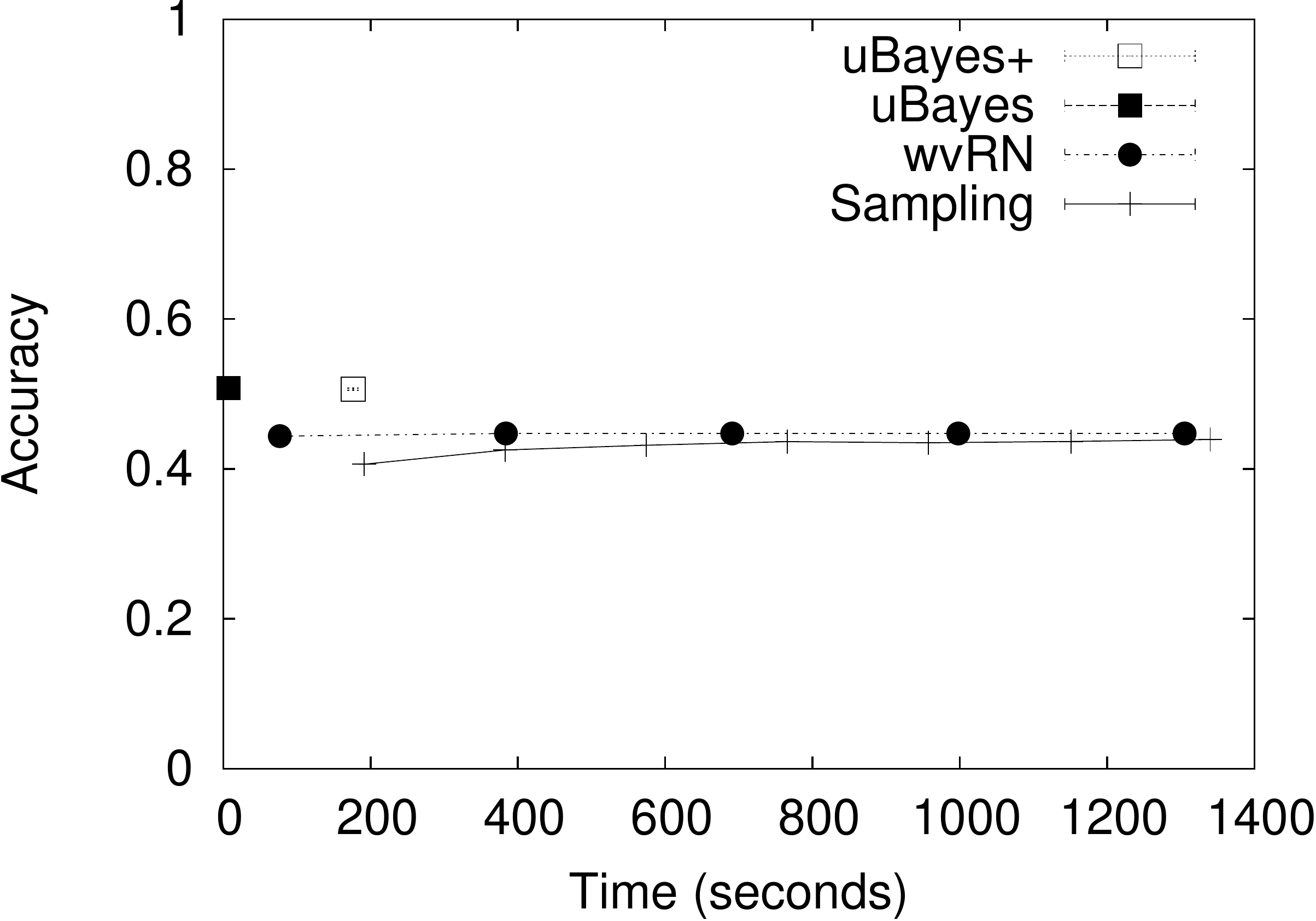}
\caption{\small\bf Accuracy  with varying execution times for {\em
uBayes}, {\em uBayes+}, {\em wvRN} and {\em Sampling} algorithms for
the {\em Patent} data set.} \label{fig:patents_timeacc}
\vspace{0.1cm}
\end{figure}

In the final set of experiments, we evaluated the accuracy of all
algorithms as a function of the  time required for algorithmic
execution by the baselines. Since the baselines tradeoff between
running time and accuracy, it is natural to include the running time
in the comparison process. In this case, we removed the constraint
that {\em wvRN} and {\em Sampling} end their processing after a
fixed amount of time or a specific number of iterations, and
examined how their accuracy changes when the number of iterations
(and consequently, processing time) increases. For reference, we
also include the {\em uBayes} and {\em uBayes+} algorithms, which
execute in a fixed amount of time. The results for the {\em DBLP}
and {\em Patent} data sets are depicted in
Figures~\ref{fig:dblp_timeacc} and \ref{fig:patents_timeacc}
respectively. The graphs show that the accuracy of {\em wvRN} and
{\em Sampling} is slightly increasing with  time, but reaches an
almost stable state after the first $10$ iterations. (In our
experiments, we stopped {\em wvRN} after $28$ iterations in the {\em
DBLP} data set and $16$ iterations in the {\em Patent} data set, and
the {\em Sampling} algorithm after $92$ iterations in the {\em DBLP}
data set and $28$ iterations in the {\em Patent} dataset).
Nevertheless, the  {\em uBayes} and {\em uBayes+}  algorithms
achieve significantly better results in a much lower running time.

%
%
%
%

\section{Conclusions}
\label{sec:conclusions}

Uncertain graphs are becoming increasingly popular in a wide variety of data domains. 
This is due to the statistical methods used to infer many networks, such as protein interaction networks and other link-prediction based methods. 
Consequently, the problem of collective classification has become particularly relevant for determining node properties in such networks. 

In this paper, we formulate the collective classification problem for uncertain graphs, and describe effective and efficient solutions for this problem.
To this effect, we describe an iterative probabilistic labeling method, based on the Bayes model, that treats uncertainty on the edges of the graph as first class citizens. 
In the proposed approach, the uncertainty probabilities of the links are used directly in the labeling process.
Furthermore, the methodology we describe allows for automatic parameter selection.

We have performed an experimental evaluation of the proposed approach using diverse, real-world datasets. 
The results show significant advantages of using such an approach for the classification process over more conventional methods, which do not directly use uncertainty probabilities.


\noindent{\bf \large Acknowledgments}\\ 
Part of this work was supported by the FP7 EU IP project KAP (grant agreement no. 260111).
Work of the second author was sponsored by the Army Research Laboratory under cooperative agreement number W911NF-09-2-0053.

\def\thebibliography#1{
  \section{References}
  \normalsize
  \list
    {[\arabic{enumi}]}
    {\settowidth\labelwidth{[#1]}
     \leftmargin\labelwidth
     \parsep 1.9pt                
     \itemsep 1.3pt               
     \advance\leftmargin\labelsep
     \usecounter{enumi}
    }
  \def\newblock{\hskip .11em plus .33em minus .07em}
  \sloppy\clubpenalty10000\widowpenalty10000
  \sfcode`\.=1000\relax
}


\end{document}